\documentclass[acmtog]{acmart}
\AtBeginDocument{%
  }

\usepackage{booktabs} 

\usepackage[ruled]{algorithm2e} 

\SetAlFnt{\small}
\SetAlCapFnt{\small}
\SetAlCapNameFnt{\small}
\SetAlCapHSkip{0pt}

\usepackage{subcaption}
\usepackage{multirow}

\usepackage{amsmath}

\DeclareMathOperator*{\argmin}{arg\,min}

\newtheorem{assumption}{Assumption}

\newcommand{\revision}[1]{\textcolor{black}{#1}}

\acmJournal{TOG}

\begin{document}
\setcopyright{acmlicensed}
\acmJournal{TOG}
\acmYear{2025} \acmVolume{44} \acmNumber{4} \acmArticle{}
\acmMonth{8}\acmDOI{10.1145/3730937}

\title{Transformer IMU Calibrator: Dynamic On-body IMU Calibration for Inertial Motion Capture}

\author{Chengxu Zuo}
\affiliation{%
 \institution{Xiamen University}
 \city{Xiamen}
 \country{China}
 }
 \email{zuochengxu@stu.xmu.edu.cn}

\author{Jiawei Huang}
\affiliation{%
 \institution{Xiamen University}
 \city{Xiamen}
 \country{China}
 }
\email{30920231154349@stu.xmu.edu.cn}

\author{Xiao Jiang}
\affiliation{%
\institution{Xiamen University}
 \city{Xiamen}
 \country{China}
 }
\email{ferster@stu.xmu.edu.cn}

\author{Yuan Yao}
\affiliation{%
 \institution{Xiamen University}
 \city{Xiamen}
 \country{China}
}
\email{furtheryao@stu.xmu.edu.cn}

\author{Xiangren Shi}
\affiliation{%
 \institution{Bournemouth University}
 \city{Bournemouth}
 \country{United Kingdom}
 }
 \email{xshi@bournemouth.ac.uk}

\author{Rui Cao}
\affiliation{%
 \institution{Xiamen University}
 \city{Xiamen}
 \country{China}
}
\email{mec2109494@xmu.edu.my}

\author{Xinyu Yi}
\affiliation{%
 \institution{Tsinghua University}
 \department{School of Software and BNRist}
 \city{Beijing}
 \country{China}}
 \email{yixy20@mails.tsinghua.edu.cn}

\author{Feng Xu}
\affiliation{%
 \institution{Tsinghua University}
 \department{School of Software and BNRist}
 \city{Beijing}
 \country{China}}
 \email{feng-xu@tsinghua.edu.cn}

\author{Shihui Guo*}
\affiliation{%
 \institution{Xiamen University}
 \city{Xiamen}
 \country{China}
}
\email{guoshihui@xmu.edu.cn}

\author{Yipeng Qin}
\affiliation{%
 \institution{Cardiff University}
 \city{Cardiff}
 \country{United Kingdom}
}
\email{qiny16@cardiff.ac.uk}

\renewcommand\shortauthors{Chengxu Zuo, Jiawei Huang, Xiao Jiang. et al}

\begin{abstract}
In this paper, we propose a novel dynamic calibration method for sparse inertial motion capture systems, which is the first to break the restrictive {\it absolute static assumption} in IMU calibration, i.e., the coordinate drift $R_{G^{'}G}$ and measurement offset $R_{BS}$ remain constant during the entire motion, thereby significantly expanding their application scenarios.
Specifically, we achieve real-time estimation of $R_{G^{'}G}$ and $R_{BS}$ under two relaxed assumptions: i) the matrices change negligibly in a short time window; ii) the human movements/IMU readings are diverse in such a time window.
Intuitively, the first assumption reduces the number of candidate matrices, and the second assumption provides diverse constraints, which greatly reduces the solution space and allows for accurate estimation of $R_{G^{'}G}$ and $R_{BS}$ from a short history of IMU readings in real time.
To achieve this, we created synthetic datasets of paired $R_{G^{'}G}$, $R_{BS}$ matrices and IMU readings, and learned their mappings using a Transformer-based model.
We also designed a calibration trigger based on the diversity of IMU readings to ensure that assumption ii) is met before applying our method.
To our knowledge, we are the first to achieve implicit IMU calibration (i.e., seamlessly putting IMUs into use without the need for an explicit calibration process), as well as the first to enable long-term and accurate motion capture using sparse IMUs. \revision{The code and dataset are available at \href{https://github.com/ZuoCX1996/TIC}{\textit{https://github.com/ZuoCX1996/TIC}}.}

\end{abstract}


%
%

\maketitle

\section{Introduction}
\label{sec:intro}
Motion capture with sparse inertial sensors has received increasing attention in recent years, given their advantage in high usability
\begin{figure}[ht]
\vspace{-2mm}
    \centering
    \includegraphics[width=0.47\textwidth]{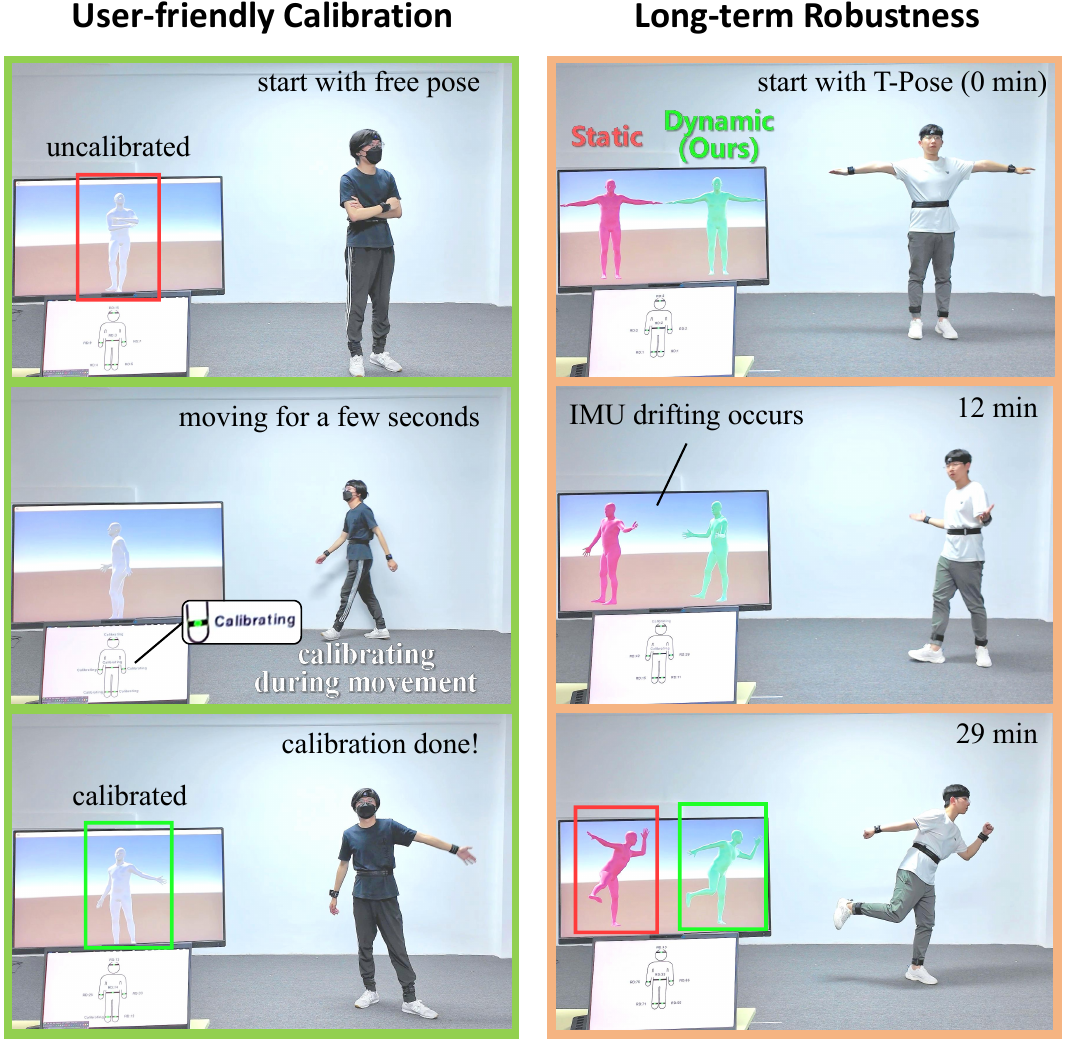}
    \vspace{-1mm}
    \caption{
    Live demonstration of our {\it dynamic} calibration method against the conventional {\it static} calibration method. Our method provides a user-friendly experience \revision{(\textbf{w/o IMU heading reset and T-Pose})} and ensures long-term robustness for inertial motion capture.}
    \label{fig:Teaser}
    \vspace{-2mm}
\end{figure}
and reduced hardware cost~\cite{huang2018deep,yi2021transpose,jiang2022transformer,yi2022physical,yi2024pnp}. These systems predict full-body skeletal rotations $\boldsymbol{\theta}$ based on partial skeletal orientations $R_{GB}\in \mathbb{R}^{3\times3}$ and their associated joint endpoint acceleration $a_{G}\in \mathbb{R}^{3\times1}$ in the global coordinate system $G$:
$
    \boldsymbol{\theta} = f(R_{GB}, a_{G})
   \label{eq:inertial_motion_capture}
$.
In practice, $R_{GB}$ and $a_{G}$ are measured by Inertial Measurement Units (IMUs) readings: rotation $R_{\rm IMU}$ and acceleration $a_{\rm IMU}$, respectively, which are defined as:
\begin{align}
\begin{split}  
    R_{\rm IMU}(t)&=R_{G^{'}G}(t)\cdot  R_{GB}(t) \cdot R_{BS}(t)\\
    a_{\rm IMU}(t)&=R_{G^{'}G}(t)\cdot  a_{G}(t)
    \label{eq:imu_partial_pose}
\end{split}
\end{align}
where $t$ denotes time;  $G^{'}$ is $G$'s offset version; $R_{G^{'}G}\in \mathbb{R}^{3\times3}$ denotes the {\it coordinate drift} caused by factors such as magnetic field interference and/or gyroscope integration error; 
$R_{BS}$ denotes the {\it measurement offset} caused by ambiguous wearing orientation of sensor, where $B$ represents the bone for which the attitude is to be measured and $S$ represents the IMU sensor.
To faithfully measure $R_{GB}$ and $a_{G}$ with $R_{\rm IMU}$ and $a_{\rm IMU}$, a {\it calibration} process is required to eliminate the effects of $R_{G^{'}G}$ and $R_{BS}$.

Traditionally, the calibration performed with an ideal but restrictive assumption that we call:
\begin{assumption}[Long-term Static Assumption]
The coordinate drift $R_{G^{'}G}(t)$ and measurement offset $R_{BS}(t)$ remain constant during the entire motion sequence (t=0, 1, 2, ..., T), i.e., 
\begin{itemize}
    \item $R_{G^{'}G}(t) = R_{G^{'}G}(0)=I$
    \item $R_{BS}(t)=R_{BS}(0)=\hat{R}_{BS}(0)$
\end{itemize}
where $\hat{R}_{BS}(0)$ is estimated using a calibration pose (e.g., T-pose). 
\label{assump:Absolute Static Assumption}
\end{assumption}
\noindent
Under this assumption, IMU calibration can {\it only} be performed at the start of motion, and it has become common practice for users to first calibrate the IMU to estimate and remove $R_{G^{'}G}(0)$ using tools provided by the sensor manufacturer before wearing the device, then to wear the device and perform a specific calibration pose to estimate and remove $R_{BS}(0)$, and finally to start motion capture.
However, such static calibration is only a {\it makeshift solution} because of the restrictive nature of the underlying Assum.~\ref{assump:Absolute Static Assumption}: 
\begin{itemize}
    \item $R_{G^{'}G}(t)$ change dynamically due to magnetic field interference and cumulative error of the gyroscope;
    \item $R_{BS}(t)$ vary over time due to accumulated changes, such as IMU placement offsets in long-term use; 
    \item Assuming $R_{BS}(0)=\hat{R}_{BS}(0)$ is suboptimal due to users' imperfect calibration poses~\cite{yi2024pnp};
\end{itemize}

These shortcomings indicate that conventional static calibration struggles to remain robust in long-term use and is sensitive to the accuracy of the calibration pose.
This results in a suboptimal user experience, limiting the adoption of inertial motion capture in applications such as gaming and sports fitness.

In this paper, we address these shortcomings by proposing a novel {\it dynamic} on-body calibration technique that operates automatically and imperceptibly during motion capture (Fig.~\ref{fig:Teaser}).
Our key insight is that although estimating $R_{G^{'}G}(t)$ and $R_{BS}(t)$ from $R_{\rm IMU}(t)$ is an inherently {\it ill-posed problem}, a feasible solution can be achieved by reducing its solution space and imposing additional constraints.
Specifically, we replace the restrictive Assum.~\ref{assump:Absolute Static Assumption} with 2 relaxed assumptions:
\begin{assumption}[Short-term Static Assumption]
Let $[t_a, t_b]$ be a short time window in a motion sequence, $t_i \in [t_a, t_b]$, we assume coordinate drift and measurement offset matrices $R_{G^{'}G}(t_i)$ and $R_{BS}(t_i)$ change negligibly in $[t_a, t_b]$, i.e.:
\begin{itemize}
    \item $R_{BS}(t_i) \approx R_{BS}(t_b)$;
    \item $R_{G^{'}G}(t_i) \approx R_{G^{'}G}(t_b)$.
\end{itemize}
\label{assump:Short-term Static Assumption}
\end{assumption}
\begin{assumption}[Short-term Diversity Assumption]
Let $t_i, t_j \in [t_a, t_b]$, $i \neq j$, $R_{\rm IMU}(t)$ is diverse in $[t_a, t_b]$, i.e.:
\begin{itemize}
    \item $R_{\rm IMU}(t_i) \neq R_{\rm IMU}(t_j)$; 
\end{itemize}
\label{assump:Short-term Diversity Assumption}
\end{assumption}
\noindent 
Then, given a short history of IMU readings $\{ R_{\rm IMU}(t_1),...,R_{\rm IMU}(t_n)\}$, $t_i \in [t_a, t_b]$ ($1\leq i \leq n$), Assum.~\ref{assump:Short-term Static Assumption} reduces its solution space from $\{R_{G^{'}G}(t_1),...,R_{G^{'}G}(t_n)\}$ and $\{R_{BS}(t_1),...,R_{BS}(t_n)\}$ to $R_{G^{'}G}(t_b)$ and $R_{BS}(t_b)$; and the inherent diversity among its elements serves as additional constraints (Assum.~\ref{assump:Short-term Diversity Assumption}).
Thanks to the arbitrary choices of $[t_a, t_b]$, our method can track dynamic changes of $R_{G^{'}G}$ and $R_{BS}$ in the entire motion sequence. 
To achieve this, we created two synthetic datasets of paired ($R_{G^{'}G}$, $R_{BS}$) and $R_{\rm IMU}$ using the AMASS~\cite{mahmood2019amass} and DIP~\cite{huang2018deep} datasets, and learned the mapping between them using a Transformer-based model. We also designed a calibration trigger based on the diversity of IMU readings to ensure that Assum.~\ref{assump:Short-term Diversity Assumption} is met before applying our method. 
Empirically, we compare our dynamic calibration with traditional static calibration methods in bone orientation and global acceleration measurement accuracy, and apply to 6 state-of-the-art sparse inertial motion capture methods, demonstrating its superiority in user-friendliness and long-term use.
In summary, our main contributions include:

\begin{itemize}
    \item Conceptually, we propose 2 fundamental assumptions (Assums.~\ref{assump:Short-term Static Assumption} and~\ref{assump:Short-term Diversity Assumption}) for dynamic calibration in sparse inertial motion capture, enabling accurate and on-the-fly estimation of IMU coordinate drift and measurement offset.
    \item Technically, we propose a practical dynamic calibration workflow including 1) a Transformer IMU Calibrator (TIC) network to estimate $R_{G^{'}G}$ and $R_{BS}$ with a short history of IMU orientations and accelerations; 2) a calibration trigger mechanism based on IMU rotation diversity to ensure effective use of TIC. To our knowledge, we are the {\it first} to achieve implicit IMU calibration, i.e., seamlessly putting IMUs into use without requiring an explicit calibration process.
    \item Additionally, we collected the first long-duration inertial motion capture dataset that explicitly incorporates IMU coordinate drift and measurement offset, providing a valuable resource for analyzing their characteristics.
\end{itemize}

\section{Related Works}
\subsection{Inertial Motion Capture}
Inertial motion capture offers advantages in portability, privacy, and resilience to challenging lighting and occlusion conditions compared to vision-based methods. 
Popular commercial IMU-based systems like Xsens~\cite{paulich2018xsens} and Noitom~\cite{noitom2017perception} use multiple wear-on IMU to capture user's motion.

Recent studies have achieved posture estimation using a sparse set (3-6) of IMUs~\cite{huang2018deep,yi2021transpose,jiang2022transformer,yi2022physical,mollyn2023imuposer,van2024diffusionposer,zhang2024dynamic}. 
TransPose~\cite{yi2021transpose} enhanced sparse IMUs motion capture by integrating multi-stage pose estimation alongside a fused global displacement estimation, integrating a module for optimizing physical dynamics in subsequent endeavors~\cite{yi2022physical}. TIP~\cite{jiang2022transformer} incorporated Transformer architecture into sparse inertial motion capture, thereby accounting for human motion in non-planar scenarios simultaneously. 
A real-time full-body posture estimation system, IMUPoser~\cite{mollyn2023imuposer} \revision{and the follow-up research MobilePoser~\cite{xu2024mobileposer}} utilize IMU data from consumer-level devices to estimate body pose and global translation. Xiao \cite{xiao2024fast} propose a hight efficiency network architecture which enabled the deployment on mobile terminals.
DynaIP~\cite{zhang2024dynamic} involves the integration of real-world motion capture data from diverse skeleton formats and introduces a novel part-based approach to enhance the robustness and accuracy of pose estimation.
DiffusionPoser~\cite{van2024diffusionposer} allows immediate use of arbirtary sensor configurations and thus optimizing these configurations for specific activities.
LIP~\cite{zuo2024loose} presents a loose-wear jacket equipped with 4 IMUs for comfortable upper-body motion tracking. 

Some emerging research explores the fusion of multiple sensor modalities to capture specific limb movements or body parts because of the high expectation to precision and flexibility, such as Ultra Inertial Poser (UIP)~\cite{armani2024ultra} and SmartPoser~\cite{devrio2023smartposer}, which introduce ultra-wideband (UWB) sensors to improve accuracy. 
More recent studies taking the advance of AR/VR applications have explored various approaches that can further sparsify input requirements to only the upper body to estimate the entire body pose by head and hand poses~\cite{Ahujia2021virtual,Du_2023_CVPR,jiang2023egoposer,jiang2022avatarposer,yang2021sparse,Zheng_2023_ICCV}. 
To address the ongoing difficulties in achieving precise joint angle and position estimations, researchers have reconsidered the use of external~\cite{pan2023fusing,Marcard_2018_ECCV} or body-worn~\cite{yi2023ego} cameras for visual-inertial tracking. 
In contrast, EM-Pose~\cite{Kaufmann_2021_ICCV} assesses the relative 3D offsets and orientations between joints through the utilization of 6–12 custom electromagnetic (EM) field-based sensors.

In all these methods, a calibration process is typically required during system initialization to acquire accurate skeleton measurement.
However, this calibration cannot effectively address the dynamic change of calibration parameters, posing a challenge in maintaining accurate pose estimation over prolonged usage.

\subsection{IMU Calibration}
With the miniaturization MEMS sensors, IMUs have been widely used for tasks such as navigation and attitude estimation. 
However, low-cost MEMS-based IMUs are usually affected by axes misalignment, bias and cross-axis sensitivities, leading to significant systematic errors in measurements~\cite{HARINDRANATH2024114001}. Traditional intrinsic calibration methods ~\cite{titterton2004strapdown, kim2004initial, syed2007new} often require expensive high-precision equipment.
\cite{tedaldi2014robust} achieve highly accurate estimation of correction parameters by placing the IMU in a set of different static positions, without the use of external equipment. \cite{li2012situ} propose a Kalman filter technique which enables the gyro triad and the accelerometer triad to calibrate each other by applying the pseudo-observations, eliminating the need for the quasi-static stays at different attitudes.

Besides the \emph{intrinsic} calibration of the IMU, calibrating the spatial \emph{extrinsic} parameters (relative translation and rotation) between the IMU and the object to which it is mounted through extrinsic calibration is equally important for IMU applications. 
In motion capture, this is commonly achieved by asking the subject to maintain a static pose, such as T-pose or N-pose, during system initialization~\cite{liu2019sensor, choe2019sensor}. Similarly, functional calibration requires the subject to perform specified movements such as hip flexion/extension and abduction/adduction ~\cite{favre2008ambulatory, favre2009functional, nazarahari2019sensor}. Since these methods rely on predefined postures and motions, their accuracy depends on how precisely the subject performs them. 
To alleviate this requirement, methods ~\cite{seel2014imu,taetz2016towards,muller2016alignment} explore the use of arbitrary motion for calibration and eliminate the need for precisely executing predefined postures or motions. 
However, these methods require attaching IMUs to adjacent joints for calibration via kinematic constraints, making them unsuitable for sparse inertial motion capture systems. 

Despite these successes, all existing calibration methods require an {\it explicit} calibration process during system initialization, which is less user-friendly and suffers from performance degradation over time.
In contrast, our work is the {\it first} to achieve implicit IMU
calibration (i.e., seamlessly putting IMUs into use without requiring an explicit calibration process), enabling user-friendly, long-term, and accurate inertial motion capture in a variety of real-world scenarios.

\section{Problem Formulation}
\subsection{IMU Calibration in Inertial Motion Capture}
As discussed in Sec.~\ref{sec:intro}, in inertial motion capture, IMU calibration aims to estimate and remove (via multiplying inverse matrix) coordinate drift and measurement offset matrices $R_{G^{'}G}$ and $R_{BS}$ from Eq.~\ref{eq:imu_partial_pose} to obtain calibrated IMU readings ${R}_{\rm cali}(t)$ and ${a}_{\rm cali}(t)$:
\begin{align}
\label{eq:imu_calibration}
\begin{split} 
    {R}_{\rm cali}(t) &=R_{G^{'}G}^{T}(t)\cdot R_{\rm IMU}(t)\cdot  R_{BS}^{T} \\
    &= R_{G^{'}G}^{T}(t)  \cdot\textbf{[}R_{G^{'}G}(t)\cdot R_{GB}(t)\cdot R_{BS}(t)\textbf{]}\cdot  R_{BS}^{T}(t)\\&= R_{GB}(t)\\
    {a}_{\rm cali}(t)&=R_{G^{'}G}^{T}(t)\cdot {a}_{\rm IMU}(t)
    =R_{G^{'}G}^{T}(t)\cdot\textbf{[}R_{G^{'}G}(t)\cdot a_{G}(t)\textbf{]}\\
    &= a_{G}(t)
\end{split}
\end{align}
where rotation matrices $R_{G^{'}G}^{T}(t)=R_{G^{'}G}^{-1}(t)$ and $R_{BS}^{T}(t)=R_{BS}^{-1}(t)$.

\begin{figure}[t]
    \centering
    \includegraphics[width=0.47\textwidth]{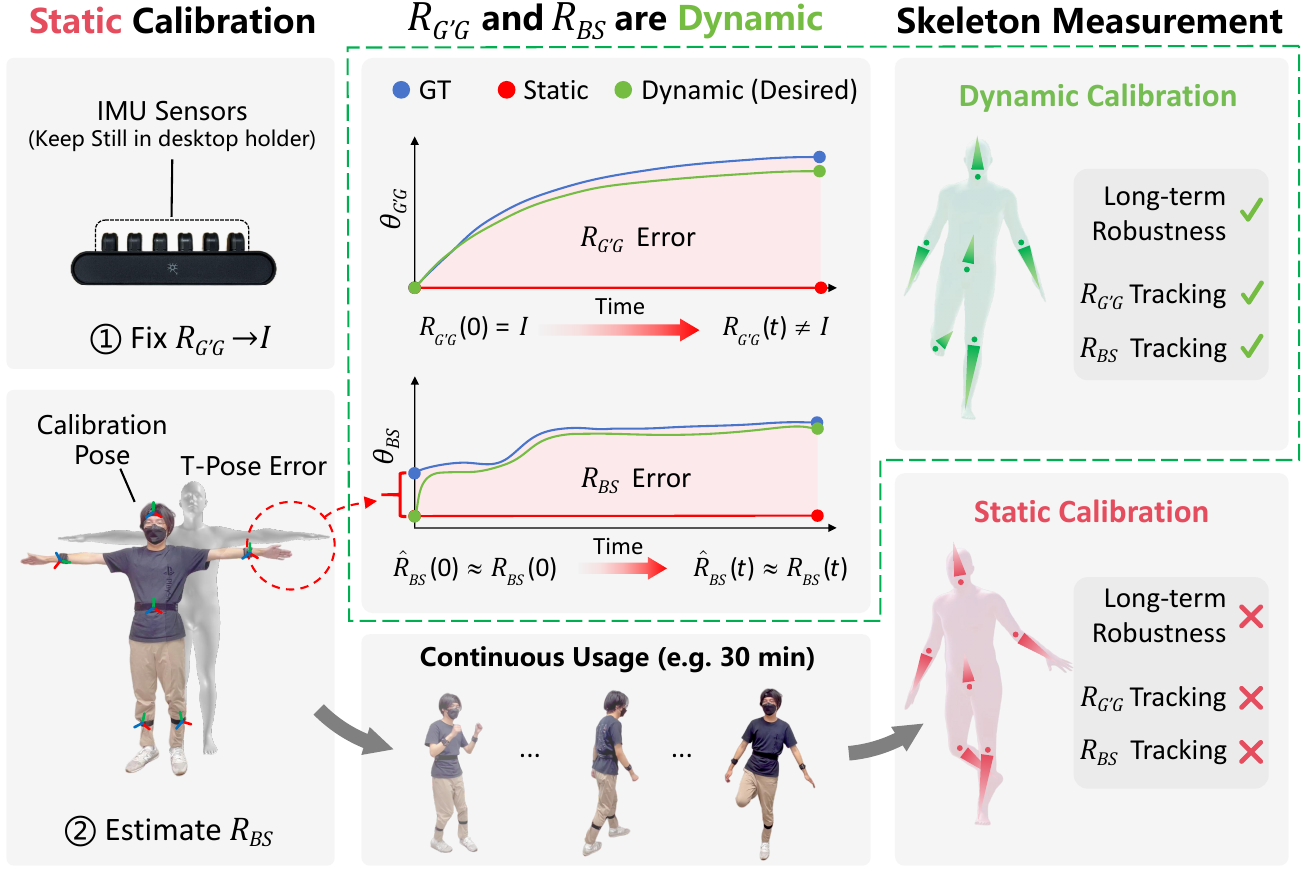}
    \caption{Motivation of our dynamic calibration. In contrast to \textcolor{red}{static} calibration that suffers from T-pose errors and the $R_{G'G}$ and $R_{BS}$ errors that increase over time, our \textcolor{green}{dynamic} calibration tracks changes of $R_{G'G}$ and $R_{BS}$ during use, ensuring long-term robustness. $\theta_{G'G}, \theta_{BS}$: rotation angles.
    }
    \label{fig:motivation}
\vspace{-4mm}
\end{figure}

\subsection{Dynamic Calibration}
The motivation for the proposed dynamic calibration arises from the intrinsic requirement to continually update calibration parameters during the motion capture (Fig.~\ref{fig:motivation}), and we define it as follows:

\paragraph{\textbf{Egocentric yaw (ego-yaw) coordinate system}}
Without loss of generality, we define the global coordinate system ${G}$ as one with zero roll and pitch rotations, while its yaw rotation is synchronized with user's body, called \textit{ego-yaw coordinate system}. 
This definition aims to eliminate the non-solvable drift component between the drifted ego-yaw and drifted world coordinate system ($R_{W^{'}G^{'}}$) (please refer to the supplementary materials).

\paragraph{\textbf{Task Definition}}
According to Assum.~\ref{assump:Short-term Static Assumption} and Assum.~\ref{assump:Short-term Diversity Assumption}, we define our dynamic calibration task as estimating:
\begin{align}
\begin{split}
    \hat{R}_{G^{'}G}(t)&=f_{d}(R_{\rm IMU}(t-n+1),...,R_{\rm IMU}(t), \\&\quad\quad\quad a_{\rm IMU}(t-n+1),...,a_{\rm IMU}(t))\\
    \hat{R}_{BS}(t)&=f_{o}(R_{\rm IMU}(t-n+1),...,R_{\rm IMU}(t), \\&\quad\quad\quad a_{\rm IMU}(t-n+1),...,a_{\rm IMU}(t))
\end{split}
\label{eq:dynamic_calibration_task}
\end{align}
where $n$ denotes the $n$-th historical IMU frame from time $t$; $f_{d}$ and $f_{o}$ are learned by:
\begin{align}
\begin{split}    
    f_{d} &= \argmin_{f} \mathbb{E}_{i,t} \mathcal{L}[\hat{R}_{G^{'}G}^{(i)}(t), {R}_{G^{'}G}^{(i)}(t)]\\
    f_{o} &= \argmin_{f} \mathbb{E}_{i,t} \mathcal{L}[\hat{R}_{BS}^{(i)}(t), {R}_{BS}^{(i)}(t)]
\end{split}
\label{eq:dynamic_calibration_model}
\end{align}
where $\mathcal{L}$ denotes a loss function; $i$ is an index of IMU sensor in training dataset consisting of paired $R_{G^{'}G}$, $R_{BS}$, and $R_{\rm IMU}$; $f$ is a hypothesis function (model) specified by the user.

\paragraph{\textbf{Acceleration Auxiliary (ACCA)}}
We introduced $a_{\rm IMU}$ as additional inputs to $f_{d}$ and $f_{o}$ because $a_{\rm IMU}$ is only influenced by ${R}_{G^{'}G}$ and is independent of $R_{BS}$ (Eq.~\ref{eq:imu_partial_pose}), which helps the model distinguish ${R}_{G^{'}G}$ and $R_{BS}$, thereby improving calibration accuracy.

\section{Method}
\begin{figure}[t]
    \centering
    \includegraphics[width=0.47\textwidth]{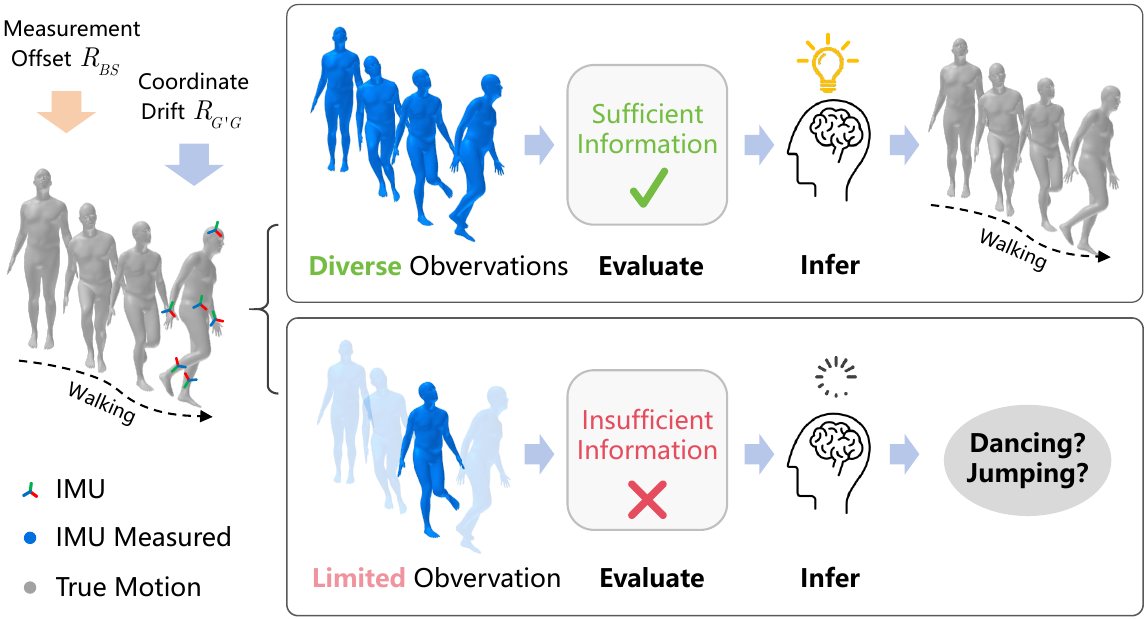}
    \caption{The rationale of our dynamic calibration. The $R_{G'G}$ and $R_{BS}$ can lead to unnatural dynamics in the captured human motion (e.g., movements that cannot maintain balance), and humans can evaluate it (our calibration trigger) and infer the original motion (our calibration).}
    \label{fig:rationale}
\vspace{-4mm}
\end{figure}

\begin{figure}[t]
    \centering
    \includegraphics[width=0.49\textwidth]{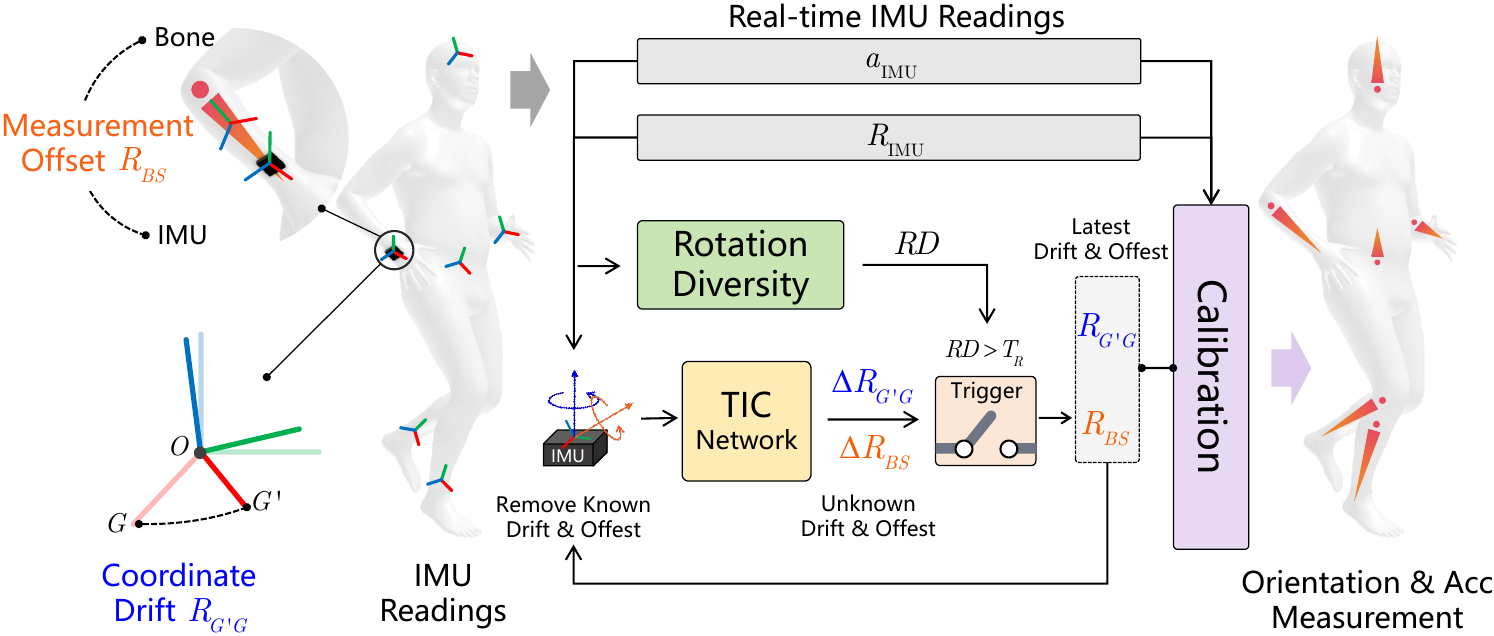}
    \caption{
    Our dynamic calibration workflow. With real-time IMU inputs, the TIC network operates at fixed short time intervals (e.g., every 2 seconds) to track the dynamic changes of $R_{G^{'}G}$ and $R_{BS}$. The updating of calibration paremeters are controlled by a Rotation Diversity ($RD$) based trigger, ensuring only reliable results that meet Assum~\ref{assump:Short-term Diversity Assumption} are used.
    }
    \label{fig:pipeline}
\end{figure}

Our dynamic calibration includes 1) \textit{TIC Network} for $R_{G'G}$ and $R_{BS}$ estimation and 2) \textit{Calibration Trigger via Rotation Diversity} ($RD$). Fig.~\ref{fig:rationale} illustrates the rationale behind this design: our calibration trigger mimics the human ability to evaluate whether a given short motion sequence contains sufficient information to determine its naturalness, while our TIC network mimics the human ability to infer the original motion through calibration.
Furthermore, we filter out unreliable results based on $RD$ to satisfy the diversity requirement in Assum.~\ref{assump:Short-term Diversity Assumption}.
Please see Fig.~\ref{fig:pipeline} for an intuitive illustration of our dynamic calibration workflow.

\subsection{TIC Network}
In this work, we instantiate the hypothesis function $f$ (Eq.~\ref{eq:dynamic_calibration_model}) with our Transformer IMU Calibrator (TIC) network.
\paragraph{\textbf{Network Architecture}}
Considering the dependence of the task on temporal information (IMU sequence), we opt to use a Transformer comprising two components:

\begin{itemize}
    \item Encoder (E): A standard Transformer encoder consisting of 3 Transformer encoder blocks for feature extraction;
    \item Transformer-Pooling-Mapping (TPM) module (x2): Each consists of a Transformer encoder block, a temporal average pooling layer, and a linear mapping layer; One for $R_{G^{'}G}$ (TPM$_{d}$), the other one for $R_{BS}$ (TPM$_{o}$).
\end{itemize}
Thus, we instantiate (Eqs.~\ref{eq:dynamic_calibration_task} and~\ref{eq:dynamic_calibration_model}) with our TIC as:
\begin{align}
\begin{split}    
   f_{d}(\cdot)&={\rm TPM}_{d}({\rm E}(\cdot))\\
   f_{o}(\cdot)&={\rm TPM}_{o}({\rm E}(\cdot))\\
\end{split}
\label{eq:TIC}
\end{align}

\paragraph{\textbf{Model Training}}
We instantiate $\mathcal{L}$ in Eq.~\ref{eq:dynamic_calibration_model} with an MSE loss and define the calibration loss $\mathcal{L}_{\text{cali}}$ as:
\begin{align}
\begin{split}
    \mathcal{L}_{\text{cali}}= &||f_{d}(\textbf{R}_{\rm IMU}^{1\rightarrow n},\textbf{a}_{\rm IMU}^{1\rightarrow n})-R_{G^{'}G}(n)||_2^2\\
    + &||f_{o}(\textbf{R}_{\rm IMU}^{1\rightarrow n},\textbf{a}_{\rm IMU}^{1\rightarrow n})-R_{BS}(n)||_2^2
    \label{eq:calibration loss}
\end{split}
\end{align}
where $\textbf{R}_{\rm IMU}^{1\rightarrow n}$ and $\textbf{a}_{\rm IMU}^{1\rightarrow n}$ are $n$ frames of input IMU orientation and acceleration readings, respectively.

\subsection{Calibration Trigger via Rotation Diversity}

As discussed in Sec.~\ref{sec:intro}, Assum.~\ref{assump:Short-term Diversity Assumption} should be met to ensure reliable  $R_{G^{'}G}$ and $R_{BS}$ estimation.
To achieve this, we propose a calibration trigger technique based on IMU rotation diversity. 
Rotation diversity is quantified by counting covered grid points in the discretized Euler angle space. We discretize the continuous Euler angle space ($\theta_x \in [-180, 180]$, $\theta_y \in [-90, 90]$, $\theta_z \in [-180, 180]$) at 15° intervals to form a $24\times12\times24$ discrete space $\mathbf{S}$. The rotation diversity $RD$ of an IMU sequence can be calculated using Algorithm~\ref{alg:rotation diversity}.
\begin{algorithm}
\caption{IMU Rotation Diversity}
\label{alg:rotation diversity}
\KwData{A sequence of IMU Rotation $\{R_{\rm IMU}(1),...,R_{\rm IMU}(n)\}$.}
\KwResult{Rotation Diversity $RD$.}
Initialize discrete Euler angle space $\mathbf{S} \in \mathbb{R}^{24 \times 12 \times 24}$ with $\mathbf{S}_{i,j,k} = 0$  $\forall i,j,k$\;
\For{$t=1$ \textbf{to} $n$}{
    $i,j,k \gets$ Calculate the coordinates of $R_{\rm IMU}(t)$ in $\mathbf{S}$\;
    $\mathbf{S}_{i,j,k} \gets \mathbf{S}_{i,j,k} + 1$\;
}
$RD \gets$ Count($\mathbf{S}_{i,j,k} > 0$);
\end{algorithm}
\subsection{Dynamic Calibration in Motion Capture}
Based on the TIC network and aforementioned rotation diversity, we have incorporated dynamic calibration into the existing inertial motion capture system (Alg.~\ref{alg:dynamic calibration}).
It is worth noting that for each IMU in the system, we independently calculate $RD$, and if the calibration is triggered, the TIC network will calculate $R_{G^{'}G}$ and $R_{BS}$ for all IMUs, but only calibrate those whose $RD$ exceeds the threshold $T_R$.
\begin{algorithm}
\caption{Dynamic Calibration in Motion Capture}
\label{alg:dynamic calibration}
\KwData{Data index $i=1$, Uncalibrated IMU orientation and acceleration readings $\{R_{\text{IMU}}(1),R_{\text{IMU}}(2),...\}$, $\{a_{\text{IMU}}(1),a_{\text{IMU}}(2),...\}$, Data Buffer $B_n$ with maximum length = $n$, Trained TIC Network $TIC$, Rotation diversity threshold $T_R$. Timing signal at intervals of \textit{t} seconds $S_{t}$. }
\KwResult{Calibrated IMU readings $\hat{R}_{\text{IMU}}$, $\hat{a}_{\text{IMU}}$.}
Initialize $R_{G^{'}G}$ and $R_{BS}$ with identity matrix $I$\;
\While{True}{
    $\hat{R}_{\text{IMU}}(i), \hat{a}_{\text{IMU}}(i) \gets$ Calibrate ${R}_{\text{IMU}}(i), {a}_{\text{IMU}}(i)$ using $R_{G^{'}G}$ and $R_{BS}$ (Eq.~\ref{eq:imu_calibration})\;
    $B_n$.append($R_{\rm IMU}^{(i)}$, $a_{\rm IMU}^{(i)}$)\;
    \If{$|B_n|==n$ and $S_{t}==\text{True}$}{
       $\textbf{R}_{\rm IMU}^{1\rightarrow n},\textbf{a}_{\rm IMU}^{1\rightarrow n}\gets$  $B_n$\;
        // \textcolor{gray}{Remove the known drift and offset.}\\
        $\textbf{R}_{\rm IMU}^{1\rightarrow n}, \textbf{a}_{\rm IMU}^{1\rightarrow n}\gets {R_{G^{'}G}^{T}} \cdot\textbf{R}_{\rm IMU}^{1\rightarrow n} \cdot R_{BS}^{T}$,  ${R_{G^{'}G}^{T}}\cdot\textbf{a}_{\rm IMU}^{1\rightarrow n}$\;
        // \textcolor{gray}{Estimating the unknown changes.}\\
        $\Delta R_{G^{'}G}$, $\Delta R_{BS} \gets$  $TIC(\textbf{R}_{\rm IMU}^{1\rightarrow n}$,  $\textbf{a}_{\rm IMU}^{1\rightarrow n})$\;
        $RD\gets$ RotationDiversity($B_n$)\;
        // \textcolor{gray}{Update drift and offset matrices.}\\
        \If{$RD>T_R$}{
            $R_{G^{'}G}\gets$ $R_{G^{'}G} \cdot \Delta R_{G^{'}G}$\;
            $R_{BS}\gets$ $\Delta R_{BS} \cdot R_{BS}$\;
        }
        $B_n$.clear()\;
    }
    i $\gets$ i + 1\;
}
\end{algorithm}

\section{Experiments}
\subsection{Training Data Synthesis}
\label{sec:synthetic_dataset}
Uncalibrated IMU data and their corresponding $R_{G^{'}G}(t)$ and $R_{BS}(t)$ data are necessary to train the TIC network. However, the cost of collecting such dataset is prohibitive because $R_{G^{'}G}(t)$ and $R_{BS}(t)$ are vary randomly during device usage, which cannot be manually controlled to collect sufficient samples.

To address this challenge, we adopted a data synthesis approach to obtain the required training data. Firstly, we need well-calibrated IMU data $\mathcal{D}_{\rm IMU}^{\rm cali}$. Similarly to works in inertial motion capture, the $\mathcal{D}_{\rm IMU}^{\rm cali}$ we used includes both synthetic data based on AMASS~\cite{mahmood2019amass} and real-world data from the DIP~\cite{huang2018deep} dataset, which helps the TIC network adapt to the characteristics of the real IMU signal.
Then, we simulated uncalibrated IMU data based on Assum.~\ref{assump:Short-term Static Assumption}.
Specifically, for each IMU data sequence (256 frames at 30Hz, 8.53s) in $\mathcal{D}_{\rm cali}$, we use random Euler angle transformations to create $R_{G^{'}G}$ and $R_{BS}$ and apply them to each frame in a sequence. These Euler angles were extensively sampled from a uniform distribution within fixed intervals (see supplementary materials) to cover all possible cases.


It is worth noting that $R_{\rm IMU}$ and $a_{\rm IMU}$ for each batch of data are synthesized on demand during model training.
Compared to using a pre-synthesized dataset of fixed/limited size, this can provide a more diverse set of $R_{G^{'}G}$ and $R_{BS}$ samples.

\subsection{Test Data Collection}

\begin{figure}[t]
    \centering
    \includegraphics[width=0.47\textwidth]{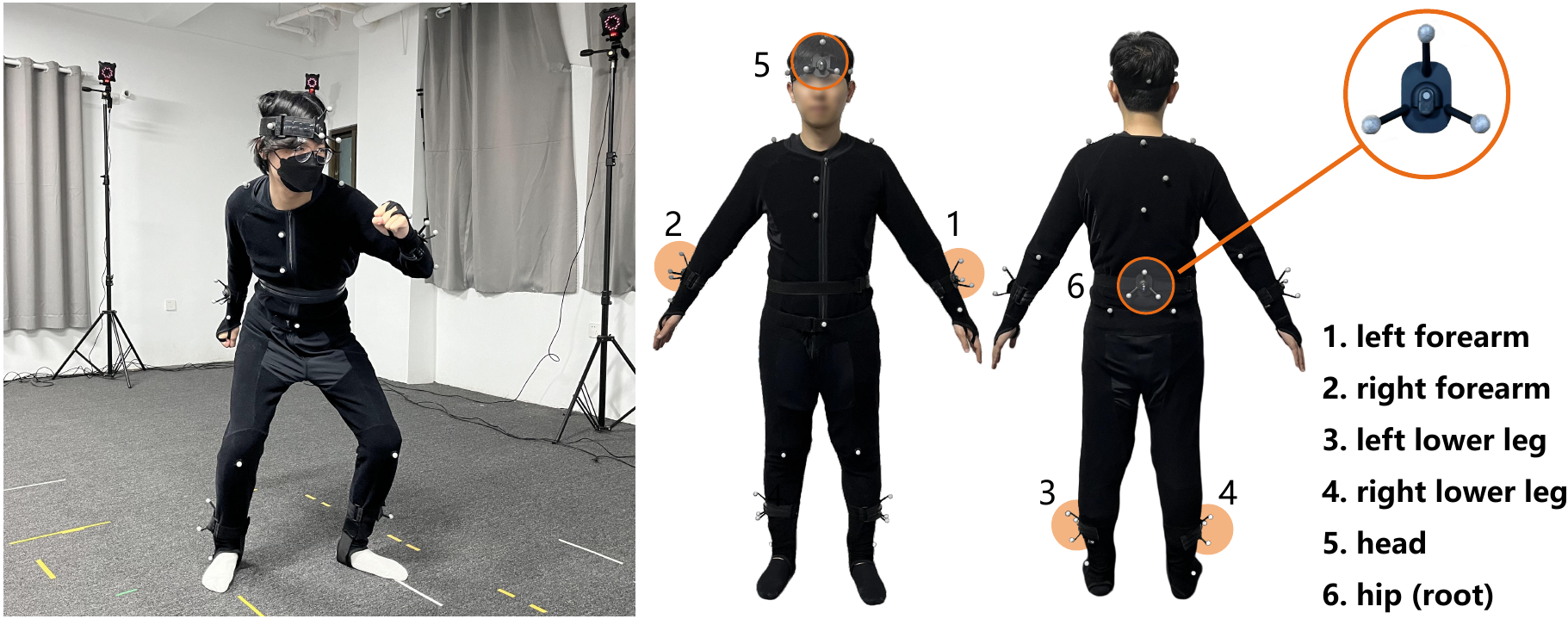}
    \caption{Our data collection system. The absolute IMU orientation and acceleration are tracked by optical markers (\textcolor{orange}{orange} circles). Body motions (skeleton orientation, global translation) and raw IMU readings are synchronously collected.}
    \label{fig:hardware}
    \vspace{-2mm}
\end{figure}

We recruited 5 volunteers (3 male, 2 female) to collect the real-world dataset. All these volunteers have experience with using inertial motion capture devices and are familiar with the process of Static Calibration. Volunteers were asked to wear both optical motion capture suits and 6 IMUs simultaneously, placed on the left forearm, right forearm, left lower leg, right lower leg, head, and hips (see Fig.~\ref{fig:hardware}). All 6 IMUs are integrated with 3 optical markers, allowing the absolute orientation and acceleration of IMUs to be captured.

We collect continuous 60-minute data at 60fps for each volunteer,
which is sufficient to capture drift signals of IMUs and ensure volunteers do not overexert themselves. In each session, volunteers sequentially perform stretching, walking/running/jumping, table tennis, aerobics, boxing, and free activity, each lasting 10 minutes. IMU data are recorded using the NOITOM Axis Lab system and calibrated through Static Calibration before system start-up. Human body pose, absolute IMU orientation and acceleration are obtained using the NOKOV motion capture system. 

Let $R_{G'S}(t)$ be the IMU orientation reading, $R_{GB}(t)$ be the captured skeleton orientation, $R_{GS}(t)$ be the absolute IMU orientation, then $R_{G^{'}G}(t)$ and $R_{BS}(t)$ can be obtained by the following formulas:

\begin{align}
\begin{split}    
R_{G^{'}G}(t)&=R_{G^{'}S}(t) \cdot R_{GS}^{T}(t) \\
R_{BS}(t)&=R_{GB}^{T}(t)\cdot R_{GS}(t)
\end{split}
\label{eq:data collection}
\end{align}
 
\begin{table}[t]
    \centering
    \caption{Dataset Summary. We collected sufficient real samples to validate the performance of the proposed dynamic calibration. $n_{\text{seq}}$: The number of IMU data sample sequences in the dataset (256 frames, 30 Hz).}
     \vspace{-2mm}
    \resizebox{0.95\linewidth}{!}{
    \begin{tabular}{l c c c c}
        \toprule
        Dataset& Purpose& $\mathcal{D}_{\text{cali}}$&$R_{G^{'}G}$ \& $R_{BS}$ &$n_{\text{seq}}$\\
        \midrule
        $\text{DS}_{\text{AMS}}$& Train& Synthesis&Synthesis &1.83M\\
        $\text{DS}_{\text{DIP}}$& Train& Real&Synthesis &114k\\
        $\text{DS}_{\text{TIC}}$& Test& Real&Real&1.04M\\
        \bottomrule
    \end{tabular}
    }
   \vspace{-4mm}
    \label{tab:Dataset Summary}
\end{table}

\subsection{Metrics of Calibration}
As shown in Eq.~\ref{eq:inertial_motion_capture}, the purpose of calibration is to obtain accurate measurements of joint orientation and global acceleration. We defined two metrics to evaluate the accuracy of calibration:
\begin{itemize}
\item \textbf{Orientation Measurement Error} (OME). The angular error between the calibrated IMU orientation and the ground-truth skeletal orientation in the ego-yaw coordinate system.
\item \textbf{Acceleration Measurement Error} (AME). The Euclidean distance between the calibrated IMU acceleration and the ground-truth acceleration (captured by NOKOV system in our work) in the ego-yaw coordinate system.
\end{itemize}

\subsection{TIC v.s. Static Calibration}
To demonstrate the advantages of applying our dynamic calibration TIC in inertial motion capture, we applied different calibration strategies on $\text{DS}_{\text{TIC}}$ and created two different datasets:
\begin{itemize}
    \item w TIC: Static calibration at system initialization, and apply TIC during usage;
    \item w/o TIC: Static calibration only at system initialization.
\end{itemize}

\begin{table}[t]
    \caption{Static Calibration vs. our TIC (dynamic) \revision{on $\text{DS}_{\text{TIC}}$}.}
    \vspace{-2mm}
    \centering
    \begin{tabular}{lrrrr}
    \toprule
        \multirow{2}{*}{Joint}&  \multicolumn{2}{c}{OME (°) $\downarrow$} & \multicolumn{2}{c}{AME ($m/s^2$) $\downarrow$}\\
        \cline{2-5}
                      &  w TIC& w/o TIC&  w TIC& w/o TIC \\
        \midrule
         left forearm &  {\bf21.40}&   63.81& {\bf1.53}&3.94\\
         right forearm&  {\bf20.56}&  68.57& {\bf1.90}&4.77\\
         left lower leg&  {\bf13.79}&  47.87& {\bf1.54}&2.17\\
         right lower leg&  {\bf12.93}& 55.23& {\bf1.42}&2.21\\
         head          &  {\bf16.56}&  57.37& {\bf0.62}&1.44\\
         hip           & {\bf6.00}& 7.03& {\bf0.80}&0.81\\
         \midrule
         Avg           & {\bf15.20}&49.98& {\bf1.30}&2.56\\
    \bottomrule
    \end{tabular}
    \label{tab:joint measurement}
\end{table}

\begin{table}[t]
 \vspace{-2mm}
    \centering
    \caption{Performance of SOTA inertial motion capture methods with / without our dynamic calibration during long-term usage \revision{(evaluated on $\text{DS}_{\text{TIC}}$)}.}
    \vspace{-2mm}
    \resizebox{\linewidth}{!}{
    \begin{tabular}{l c c c c}
        \toprule
        \multirow{2}{*}{\textbf{Method}}  & \multicolumn{4}{c}{Pose error metrics with / without dynamic calibration}\\
        \cline{2-5}&Angular (°)&Positional (cm)&SIP (°)& Mesh (cm)\\
        \midrule
        \textbf{DIP}& \textbf{19.31}/37.18& \textbf{9.36}/13.25& \textbf{20.89}/32.38& \textbf{10.97}/17.22\\
        \textbf{TransPose}& \textbf{17.90}/36.88& \textbf{8.43}/12.72& \textbf{18.98}/32.06& \textbf{10.33}/16.11\\
        \textbf{TIP}& \textbf{16.50}/37.56& \textbf{7.28}/13.41& \textbf{17.28}/33.05& \textbf{8.92}/16.86\\
        \textbf{PIP}& \textbf{16.21}/32.39& \textbf{7.53}/12.02& \textbf{15.77}/29.86& \textbf{9.30}/14.84\\
 \textbf{DynaIP}& \textbf{16.71}/35.16& \textbf{7.35}/12.14& \textbf{16.56}/30.14&\textbf{9.29}/15.02\\
        \textbf{PNP}& \textbf{15.52}/30.60& \textbf{7.20}/12.68&  \textbf{14.18}/25.13& \textbf{8.84}/15.26\\
        \bottomrule
    \end{tabular}
    }
   \vspace{-4mm}
    \label{tab:comp SOTA ang}
\end{table}

Table~\ref{tab:joint measurement} shows the OME and AME on these two datasets, indicating that dynamic calibration based on TIC provides more accurate skeleton motion measurements. It can be observed that the skeletal measurement error obtained using Static Calibration for the root node (hip) is lower. This is because in the ego-yaw coordinate frame, the yaw rotation is defined by the yaw rotation of the root node, so its pose measurement error is not affected by IMU drifting.

With dynamic updating of $R_{G^{'}G}$ and $R_{BS}$, TIC can significantly improve the robustness of inertial motion capture systems during long-term use. We used six state-of-the-art methods, DIP~\cite{huang2018deep}, Transpose~\cite{yi2021transpose}, TIP~\cite{jiang2022transformer}, PIP~\cite{yi2022physical}, DynaIP\cite{zhang2024dynamic} and PNP~\cite{yi2024pnp}, to test the effectiveness of TIC. The error metrics used include:
\begin{itemize}
\item \text{Angular Error}. The global rotation error of all joints;

\item \text{Positional Error}. The joint position error of all joints;

\item \text{SIP Error}. The global rotation error of hips and shoulders;

\item \text{Mesh Error}. The vertex error of the posed SMPL meshes.
\end{itemize}
Both metrics were calculated with the root joint (Hip) aligned.

Table~\ref{tab:comp SOTA ang} shows that pose estimation metrics with TIC are significantly lower than those without TIC. This implies that unreliable skeleton measurements caused by the calibration parameters changing severely affect pose estimation, and TIC effectively addresses this issue. In particular, we found that PNP achieved the optimal result, which is attributed to the PNP includes calibration errors simulation caused by a small volume of \( R_{G^{'}G} \) and \( R_{BS} \) in the training data, thus achieve better adaption to imperfect calibration.

\begin{figure}[t]
 \vspace{-1mm}
    \centering
    \includegraphics[width=0.98\linewidth]{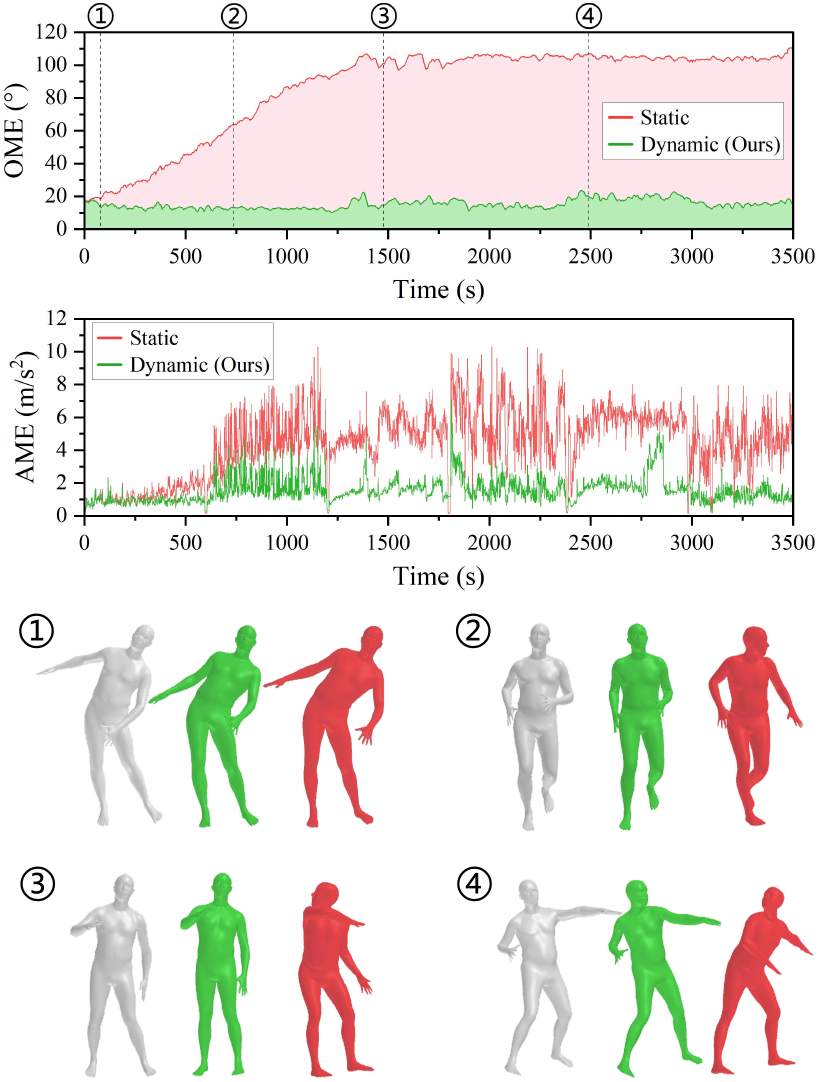}
    \caption{Top: Static Calibration (PNP) vs. our Dynamic Calibration (TIC) on joint orientation measurement error over an extended period. Bottom: visualization of predicted poses at the four time points highlighted in Top.}
    \label{fig:pose visualization}
    \vspace{-2mm}
\end{figure}

In Fig.~\ref{fig:pose visualization} we visualize both OME and the corresponding predicted poses at different time points.
The comparison demonstrates that, with only static calibration, the OME gradually increases as a result of the change of calibration parameters, leading to incorrect pose estimation. 
In contrast, thanks to the tracking of calibration parameters change, our dynamic calibration ensures accurate skeleton orientation and acceleration measurement over long durations, thereby maintaining robust pose estimation.

\subsection{Ablation Study}

To assess the necessity of the rotation diversity ($RD$) trigger and the Acceleration Auxiliary (ACCA), we recorded the error metrics of dynamic calibration after their removal. The results presented in Table~\ref{tab:ablation} justify our claim that the effectiveness of using TIC relies on Assum.~\ref{assump:Short-term Diversity Assumption}, whose fulfillment is ensured by the rotation diversity ($RD$) trigger (Case 1 vs. case 2). Furthermore, as expected, the ACCA has a significant impact on the accuracy of $R_{G^{'}G}$. The removal of ACCA resulted in an increase of 85.4\% in the $R_{G^{'}G}$ error (Case 1 vs. Case 3), which consequently led to a rise in AME.

\revision{Additionally, Case 5 presents the ablation results of $\text{DS}_{\text{DIP}}$ fine-tuning, demonstrating the benefits of real-world IMU data in TIC network training}.

\begin{figure}[t]
    \centering
    \includegraphics[width=0.49\textwidth]{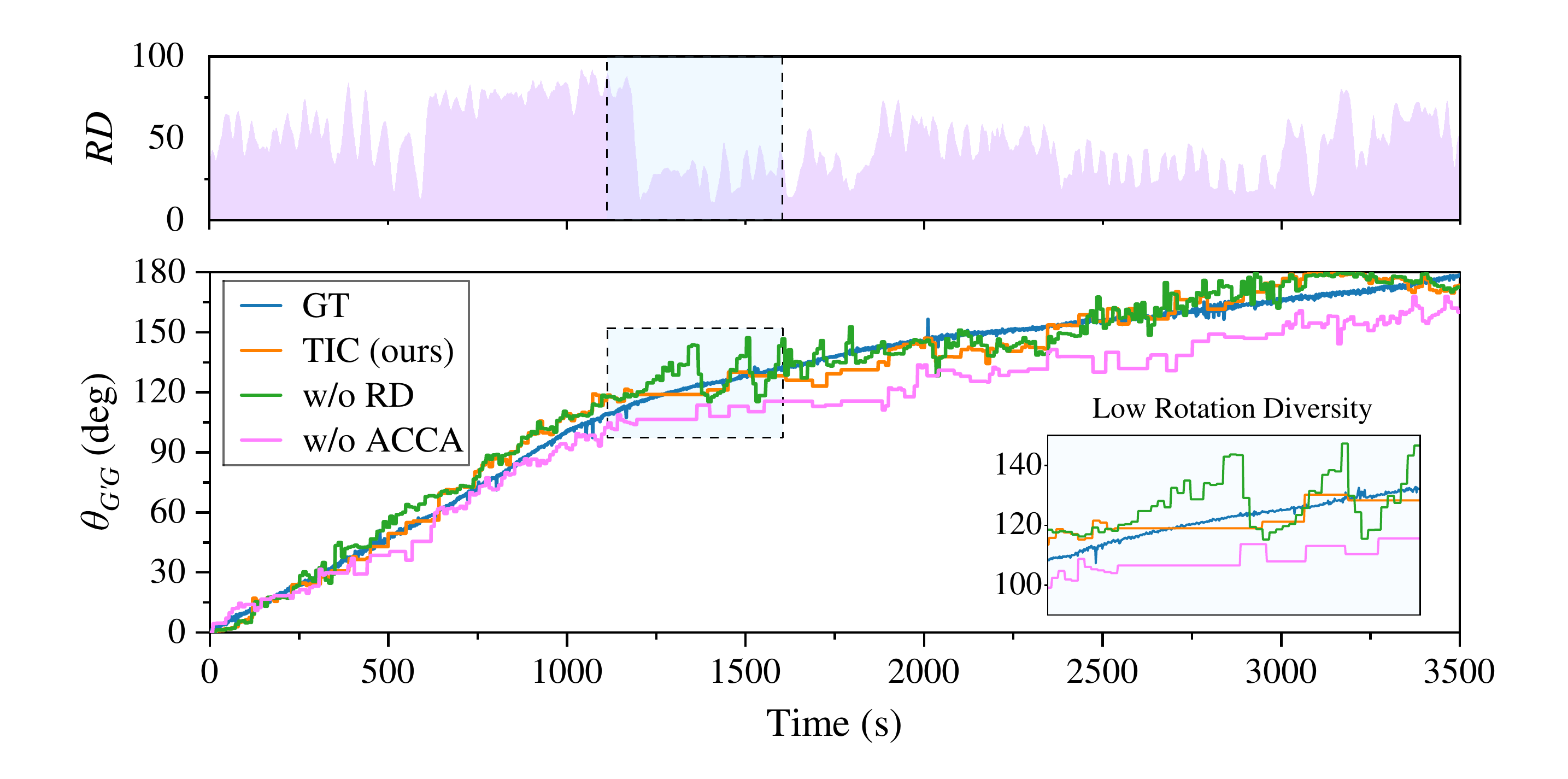}
    \caption{Qualitative evaluation on of Rotation Diversity ($RD$) Trigger and ACCA in $R_{G^{'}G}$ tracking (subject 1, right lower leg IMU in $\text{DS}_{\text{TIC}}$). $\theta_{G^{'}G}$: angle of coordinate drift $R_{G^{'}G}$ in degree; GT: ground truth coordinate drift in ego-yaw frame.}
    \label{fig:ablation}
     \vspace{-2mm}
\end{figure}

Fig.~\ref{fig:ablation} provides a more intuitive illustration of the impact of $RD$ Trigger and ACCA. It can be observed that after removing the $RD$ Trigger, there is a significant error and oscillation in \( R_{G'G} \)  tracking under low $RD$ conditions. This is due to the unreliable outputs not being filtered and applied to $ R_{G^{'}G}$ updating. On the other hand, after removing ACCA, even with $RD$ Trigger, the tracking accuracy of $ R_{G^{'}G}$ still suffers a noticeable decline. This reflects the important role of ACCA in accurate $ R_{G^{'}G}$ estimation. 

\begin{table}[t]
    \caption{Ablation Study on Rotation Diversity ($RD$) trigger, the acceleration auxiliary (ACCA) \revision{and $\text{DS}_{\text{DIP}}$ finetuning}. $R_{G^{'}G}/R_{BS}$ Err: Regression error of $R_{G^{'}G}/R_{BS}$ (°).}
    \centering
    \begin{tabular}{cccrrrr}
    \toprule
          Case&$RD$ &ACCA& OME &AME  & $R_{G^{'}G}$ Err&$R_{BS}$ Err\\
         \midrule
            1&+&+&  15.20&1.30& 9.18&15.28\\
            2&-&+&  15.48&1.32& 9.27&15.56\\
            3&+&-&  16.35&1.85& 15.22&16.53\\
            4&-&-&  16.86&1.91& 16.21&16.90\\
           \midrule
           \revision{5}&\multicolumn{2}{c}{\revision{w/o $\text{DS}_{\text{DIP}}$}}&  \revision{16.45}&\revision{1.30}& \revision{9.21}&\revision{16.38}\\
           \bottomrule
    \end{tabular}
    \label{tab:ablation}
\end{table}

\subsection{Error Analysis}
\label{sec:err ana}
\begin{figure}[t]
    \centering
    \includegraphics[width=0.98\linewidth]{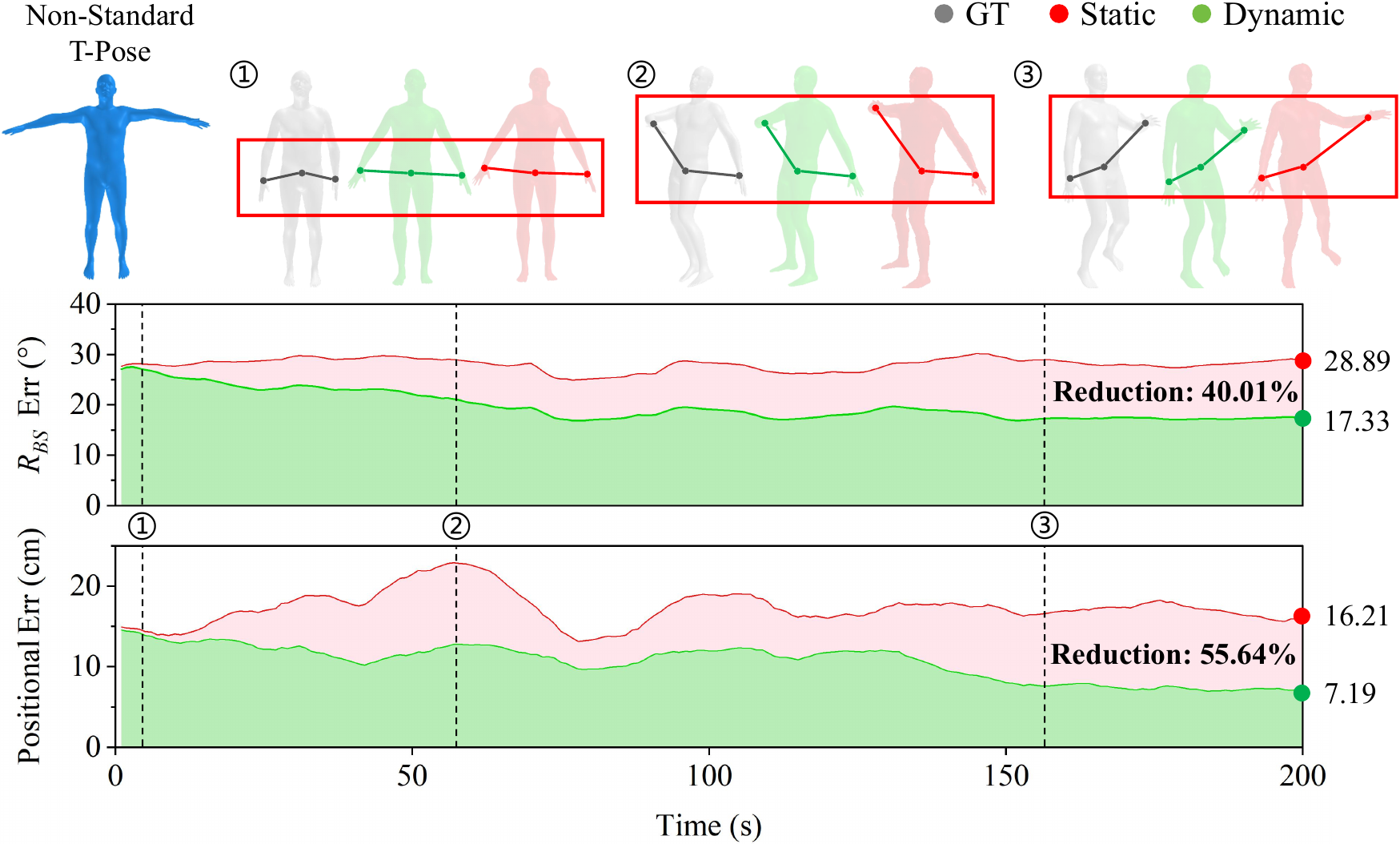}
    \caption{Qualitative evaluation on $R_{BS}$ error. The visualized $R_{BS}$ error and positional error are the average values of left and right forearm. The examples are from subject 5 in $\text{DS}_{\text{TIC}}$.}
    \label{fig:err ana}
     \vspace{-4mm}
\end{figure}

In this section, we analyze the errors in dynamic calibration. As shown in Table~\ref{tab:regression err}, the \( R_{BS} \) error is significantly higher than \( R_{G^{'}G} \), particularly for the IMUs located on the left and right forearms, which leads to higher OME. For further investigation, we visualized the \( R_{BS} \) error and positional error (calculated via the predicted human pose) for two forearm joints over the first 200 seconds of motion capture. As shown in Fig.~\ref{fig:err ana}, our dynamic calibration reduces both \( R_{BS} \) error and positional error caused by non-standard calibration poses. Notably, compared to static calibration, the reduction in positional error using dynamic calibration is significantly greater (55.64\% vs. 40.01\%), ensuring accurate joint position estimation. This indicates that our dynamic calibration more effectively tracks the rotation components affecting joint position in \( R_{BS} \), while disregarding part of rotation components unrelated to joint position (e.g., axial rotation of the forearm).

\begin{table}[t]
    \caption{Error of $R_{G^{'}G}$ and $R_{BS}$ and corresponding OME in $\text{DS}_\text{TIC}$.}
    \centering
    \begin{tabular}{lccc}
    \toprule
        IMU location&  $R_{G^{'}G}$ Err (°)&$R_{BS}$ Err (°) &OME (°)\\
        \midrule
         left forearm  &  10.49& 21.41&21.40\\
         right forearm &  15.44& 17.33&20.56\\
         left lower leg &  6.66& 14.33&13.79\\
         right lower leg &  10.19& 14.32&12.93\\
         head            &  9.94& 16.53&16.56\\
         hip             &  2.34& 7.76&6.00\\
    \bottomrule
    \end{tabular}
    \label{tab:regression err}
\end{table}

\begin{figure}[t]
    \centering
    \includegraphics[width=0.40\textwidth]{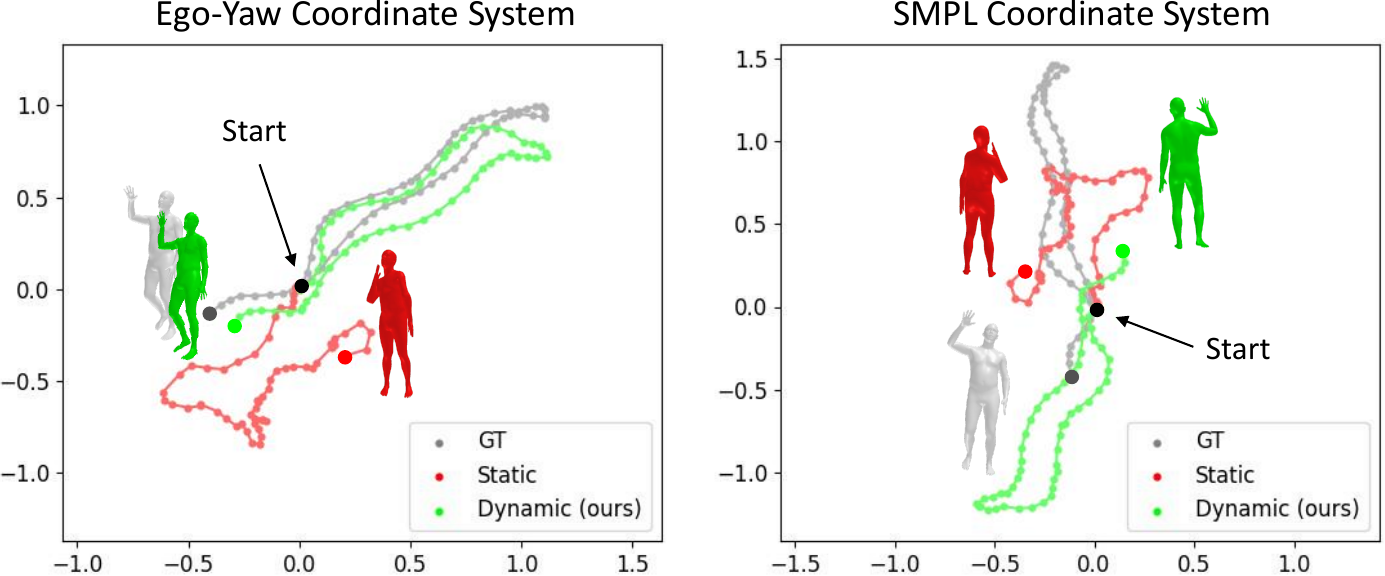}
    \caption{\revision{Visualization of global translation tracking on $\text{DS}_{\text{TIC}}$ in the SMPL frame and ego-yaw frame, unit: m.}}
    \label{fig:trans comp}
\end{figure}

\revision{
\subsection{Evaluation on Global Translation}
Our dynamic calibration significantly enhances the accuracy of translation estimation over extended periods. As shown in Table~\ref{tab:trans err}, the translation error using dynamic calibration (TIC) in the ego-yaw coordinate system is markedly lower than when it is not used. This improvement is attributed to the fact that translation estimation relies on reliable pose estimation results for forward kinematics calculations~\cite{yi2021transpose, yi2022physical, yi2024pnp}, which TIC effectively ensures. However, since TIC can only solve IMU drift within the ego yaw coordinate system, it cannot correct the drift outside the ego yaw coordinate system (such as the drift of the root IMU). Consequently, for translation in a fixed global coordinate system (e.g., the SMPL coordinate system), accuracy of translation may not be guarantee (see Fig.~\ref{fig:trans comp}).
}
\begin{table}
\caption{\revision{Translation error under different coordinate frame with / without our dynamic calibration TIC. The translation is captured using PIP~\cite{yi2022physical}. 1/2/5/10s: duration of time window. Ego-Yaw: fit the translation to the ego-yaw coordinate system of the first sample within the time window.}}
    \centering
    \begin{tabular}{cccccc}
    \toprule
 \multirow{2}{*}{Coordinate System}& \multirow{2}{*}{TIC}& \multicolumn{4}{c}{Translation Error (cm)}\\
    \cline{3-6}
         &&  1s&  2s&  5s& 10s\\
         \midrule
          \multirow{2}{*}{Ego-Yaw} &+&  \textbf{8.38}&  \textbf{12.46}&  \textbf{22.24}& \textbf{36.43}\\
            &-&  15.31&  23.16&  34.06& 44.96\\
        \midrule
          \multirow{2}{*}{SMPL}&+&  18.03&  26.79&  38.54& 49.50\\
          &-&  14.09&  19.84&  29.97& 41.69\\
        \bottomrule
    \end{tabular}
    \label{tab:trans err}
\end{table}

\revision{
\subsection{Evaluation on Consumer-grade IMU}
As shown in Table~\ref{tab:imuposer}, we also verified our dynamic calibration TIC on the IMUPoser dataset~\cite{mollyn2023imuposer}, in which data were collected from 5 consumer-grade IMUs integrated in smartphones, smartwatches, and earbuds. These IMUs were located on head, wrists and front pockets of pants, with a sensor layout different from that in our work. The results demonstrated that TIC significantly reduced OME in the IMUPoser dataset, which highlights the potential of TIC for consumer-grade inertial motion capture systems.
}
\begin{table}[h]
    \centering  
    \caption{\revision{Results on the IMUPoser Dataset.}}  
    \begin{tabular}{lrr}  
    \toprule  
    \multirow{2}{*}{IMU Location} & \multicolumn{2}{c}{OME (°)}\\
    \cline{2-3}
    & with TIC & without TIC \\
    \midrule  
    left wrist & \textbf{20.74} & 29.42 \\
    right wrist & \textbf{21.95} & 33.11 \\
    left front Pocket & \textbf{16.93} & 19.68 \\
    right front Pocket (ego-yaw) & 10.64 & \textbf{9.18} \\
    head & \textbf{24.51} & 26.66 \\
    \midrule  
    Avg & \textbf{18.95} & 23.61 \\
    \bottomrule   
    \end{tabular}
    \label{tab:imuposer}
\end{table}  

\section{Limitations}
\begin{figure}[h]
    \centering
\includegraphics[width=0.45\textwidth]{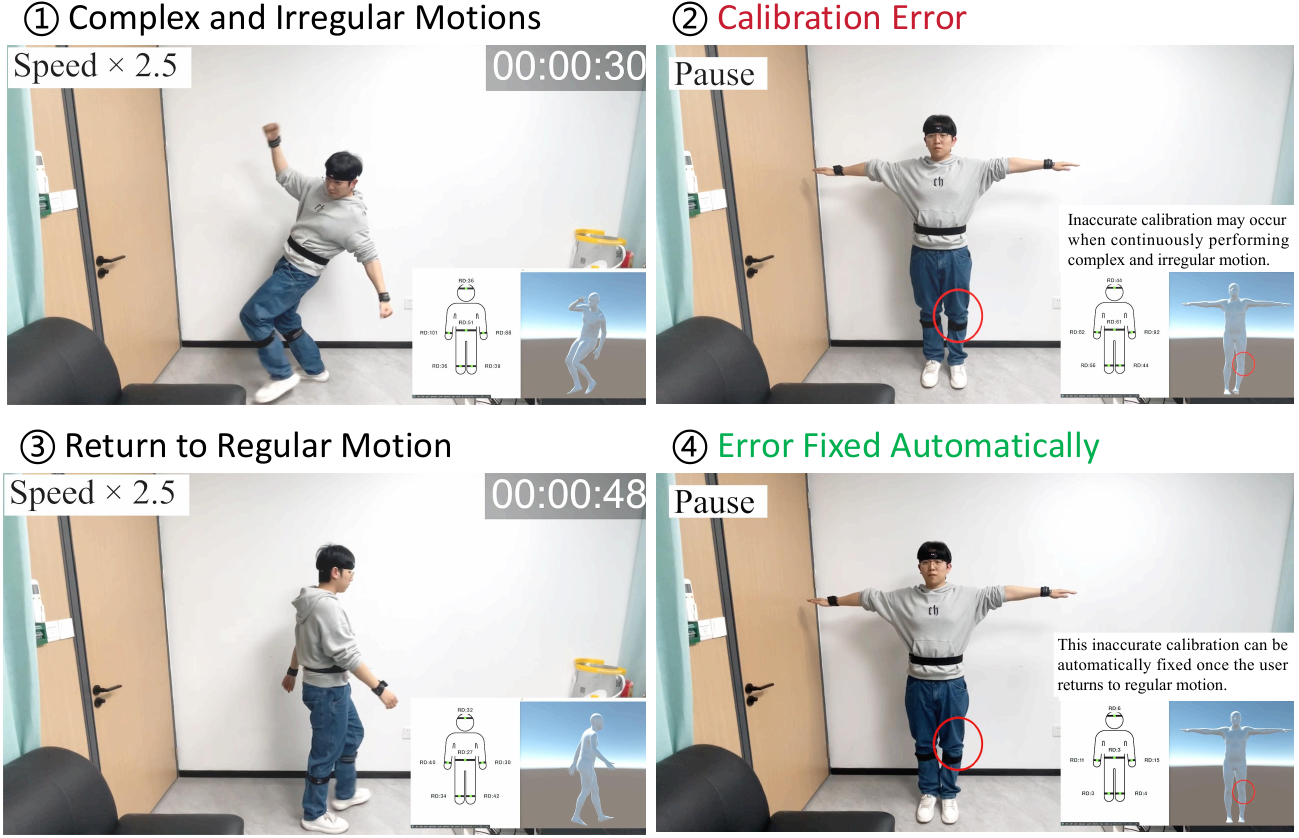}
    \caption{\revision{Evaluation on complex and irregular motions in the demo video. We observed that such scenarios could lead to inaccurate calibration, as these types of motions are underrepresented in existing motion datasets. Fortunately, our approach can automatically recover and correct the calibration once the user resumes regular motion.}}
    \label{fig:irregular demo}
\end{figure}

Since our dynamic calibration is based on Assum.~\ref{assump:Short-term Static Assumption} and Assum.~\ref{assump:Short-term Diversity Assumption}, it may fail under the following conditions: i) large and sudden changes in $R_{G^{'}G}$ and $R_{BS}$ (e.g., extreme environmental changes or sensor hardware inconsistencies); ii) low-activity scenarios, such as office work, watching TV, etc, where the rotation diversity may not meet the requirement for effective calibration; iii) we only consider the coordinate drift in the ego-yaw frame and do not support the correction of global yaw drift (Fig.~\ref{fig:trans comp}); \revision{iv) irregular motions may lead to incorrect calibration (Fig.~\ref{fig:irregular demo})}.

\section{Conclusion}
This work presented a novel dynamic calibration method for sparse inertial motion capture systems that breaks the restrictive absolute static assumption in traditional IMU calibration. 
Our key innovations include: i) real-time calibration parameters estimation under two relaxed assumptions that they change negligibly over short windows and human movements provide diverse IMU readings in that window, and ii) creating a Transformer-based model trained on synthetic data to learn the mapping from IMU readings to calibration parameters. 
Our work represents the first implicit IMU calibration technique that integrates calibration into regular usage without requiring an explicit calibration process. The promising results demonstrate the significant potential of our dynamic calibration framework in extending the capture duration and expanding the applications of inertial motion capture.

\begin{acks}
This work is supported by National Natural Science Foundation of China (62472364, 62072383), the Public Technology Service Platform Project of Xiamen City (No.3502Z20231043), Xiaomi Young Talents Program / Xiaomi Foundation and the Fundamental Research Funds for the Central Universities (20720240058). This work is partially supported by Royal Society (IEC \textbackslash NSFC \textbackslash211022). Shihui Guo is the corresponding author.
\end{acks}

\bibliographystyle{ACM-Reference-Format}
\bibliography{main}

\appendix

\section{Drift \& Offset Simulation}
\paragraph{\textbf{Orientation}}
We simulate the impact of coordinate drift and measurement offset on orientation measurement based on Assum.2 and Eq.2 in the main paper. As shown in Table~\ref{tab:random_sampling}, the $R_{G^{'}G}$ and $R_{BS}$ used are obtained by randomly sampling from a uniform distribution within a specific range. For $R_{BS}$, we set a distribution range of 45 degrees equally for rotations around the x, y, and z axes, as $R_{BS}$ is influenced by calibration pose errors and can occur around any rotation axis. For $R_{G^{'}G}$, we distinguish two cases: 1) The $R_{G^{'}G}$ for the non-root IMU is primarily set to rotate around the yaw (the y-axis in the SMPL frame), aligning with the characteristics of actual drifting; 2) The yaw rotation of the $R_{G^{'}G}$ for the root IMU is set to 0, as the root IMU measures the root joint of the human body, and its yaw rotation always be 0 in the ego-yaw coordinate system, i.e.,

\begin{align}
\begin{split}    
 Yaw(R_{G^{'}G}^{(root)} \cdot R_{GB_{root}})=Yaw(R_{GB_{root}})=0
\end{split}
\label{eq:ego-yaw root drift}
\end{align}
where $G$ represents the ego-yaw coordinate system, $G'$ represents the drifted $G$, and $B_{root}$ refers to the root bone, $R_{G'G}^{(root)}$ is the drifting applied to the root IMU. To ensure that the formula holds, the yaw rotation of $R_{G'G}^{(root)}$ must be 0.

\begin{table}[ht]
\centering
\caption{Distribution of random $R_{G^{'}G}(t)$ and $R_{BS}(t)$ samples. $U(a, b)$: a uniform distribution between a and b (inclusive of a and b).}
\begin{tabular}{lccc}
\hline
\multirow{2}{*}{\textbf{Rotation Matrix}} & \multicolumn{3}{c}{\textbf{Distribution}} \\
\cline{2-4}
 & $\Theta_x$ & $\Theta_y$ & $\Theta_z$ \\
\hline
$R_{BS}$ & $U(-45, 45)$ & $U(-45, 45)$ & $U(-45, 45)$ \\
$R_{G^{'}G}$ (root)& $U(-20, 20)$ & $U(0, 0)$ & $U(-20, 20)$ \\
$R_{G^{'}G}$ (non-root)& $U(-20, 20)$ & $U(-60, 60)$ & $U(-20, 20)$ \\
\hline
\end{tabular}
\label{tab:random_sampling}
\end{table}

\paragraph{\textbf{Hardware Level Acceleration}}
Here we illustrate our Hardware Level Acceleration simulation under coordinate drift. The global acceleration measurement of a real IMU is a \textbf{corrected value} obtained by manually removing the influence of Gravitational Acceleration (GA). The hardware level acceleration $ \tilde{a}_{G}$ can be expressed as follows:

\begin{align}
\begin{split}    
  \tilde{a}_{G}&=R_{GS}\cdot (a_{S}-g_{S})+g_{\text{bias}}\\
  &=a_{G}-g_{G}+g_{\text{bias}}
\end{split}
\label{eq:real acc drift}
\end{align}
where $a_S$ and $g_S$ are the linear and gravitational acceleration in the sensor coordinate system, respectively. The $g_{\text{bias}}$ is typically set to accurate gravitational acceleration in the global coordinate system $G$. 
In the hardware level, the coordinate drift $R_{G^{'}G}$ can lead to inaccurate GA removal, which we refer to as \textbf{GA leakage} in accelerations measurement (Fig.~\ref{fig:gravity leakage}). 
Based on Eq.2, the influence of coordinate drift on $\tilde{a}_{G}$ can be modeled as follow:

\begin{figure}[h]
    \centering
\includegraphics[width=0.45\textwidth]{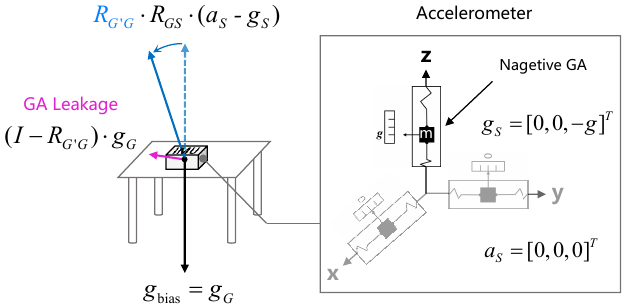}
    \caption{Illustration of gravitational acceleration (GA) leakage in hardware level acceleration measurement. The accelerometer inherently includes a negative GA measurement. The $R_{G^{'}G}$ can lead to inaccurate GA removal.}
    \label{fig:gravity leakage}
\end{figure}

\begin{align}
\begin{split}
\tilde{a}_{G^{'}}&=R_{G^{'}G}\cdot (a_{G}-g_{G})+g_{\text{bias}}\\
&=R_{G^{'}G}\cdot a_{G}-R_{G^{'}G}\cdot g_{G}+g_{G}\\
&=R_{G^{'}G}\cdot a_{G}+(I-R_{G^{'}G})\cdot g_{G}
\end{split}
\label{eq:real acc drift}
\end{align}

The term $(I - R_{GG}) \cdot g_{G}$ represents the GA leakage. During the simulation, we set $g_{\text{bias}}$ to $[0, -9.80665, 0]^T$, which accurately represents the gravitational acceleration in the ego-yaw coordinate system of the SMPL body.

\section{IMU Drifting Analysis}
A direct observation of IMU drifting is demonstrated in the supplementary video (Fig~\ref{fig:imu still}).
All six IMUs are placed in a compact holder, which ideally should maintain the IMUs to be fixed at the same orientation.
A noticeable drift is visualized when monitoring the IMU orientation in real time.
All six IMUs demonstrate different extents of drifting.

\begin{figure}[h]
    \centering
    \includegraphics[width=0.49\textwidth]{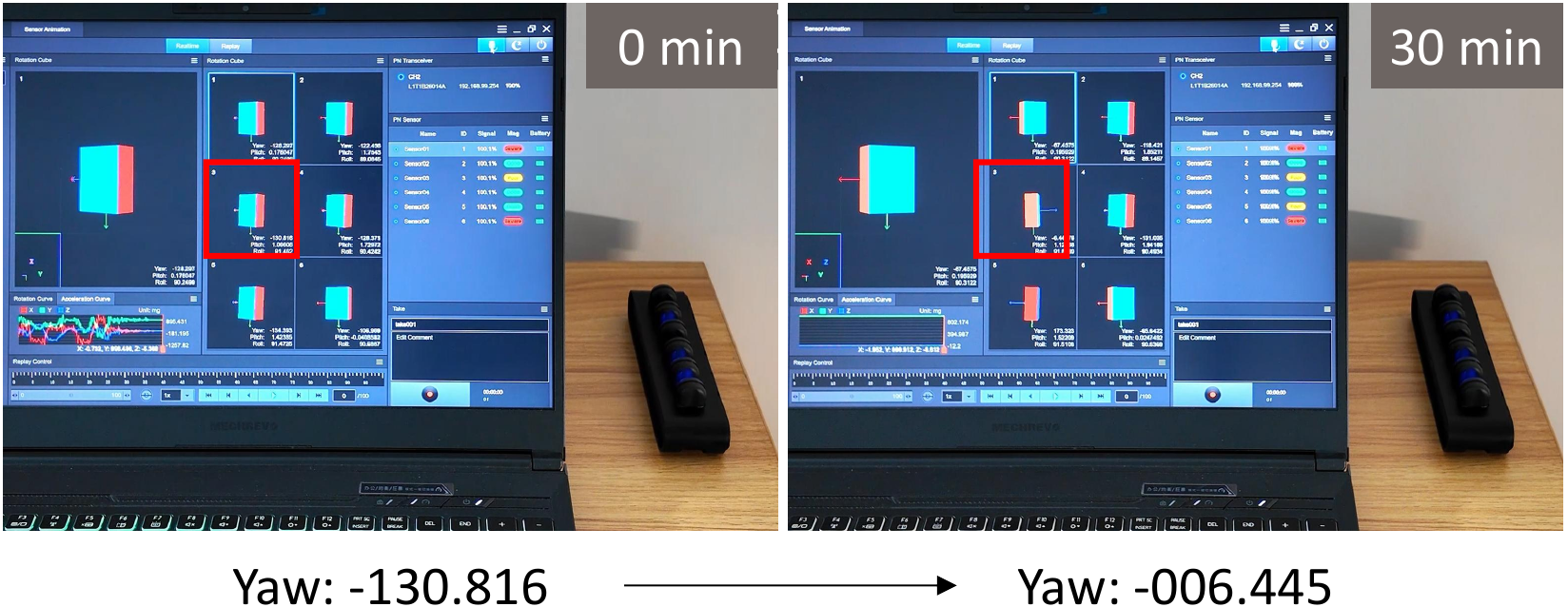}
    \caption{Visualization of IMU drifting in the scenario of placing them in a desktop holder. In the 30-minute observation, we recorded a maximum yaw drifting of 0.07 deg/s. }
    \label{fig:imu still}
\end{figure}

\begin{figure}[h]
    \centering
    \includegraphics[width=0.49\textwidth]{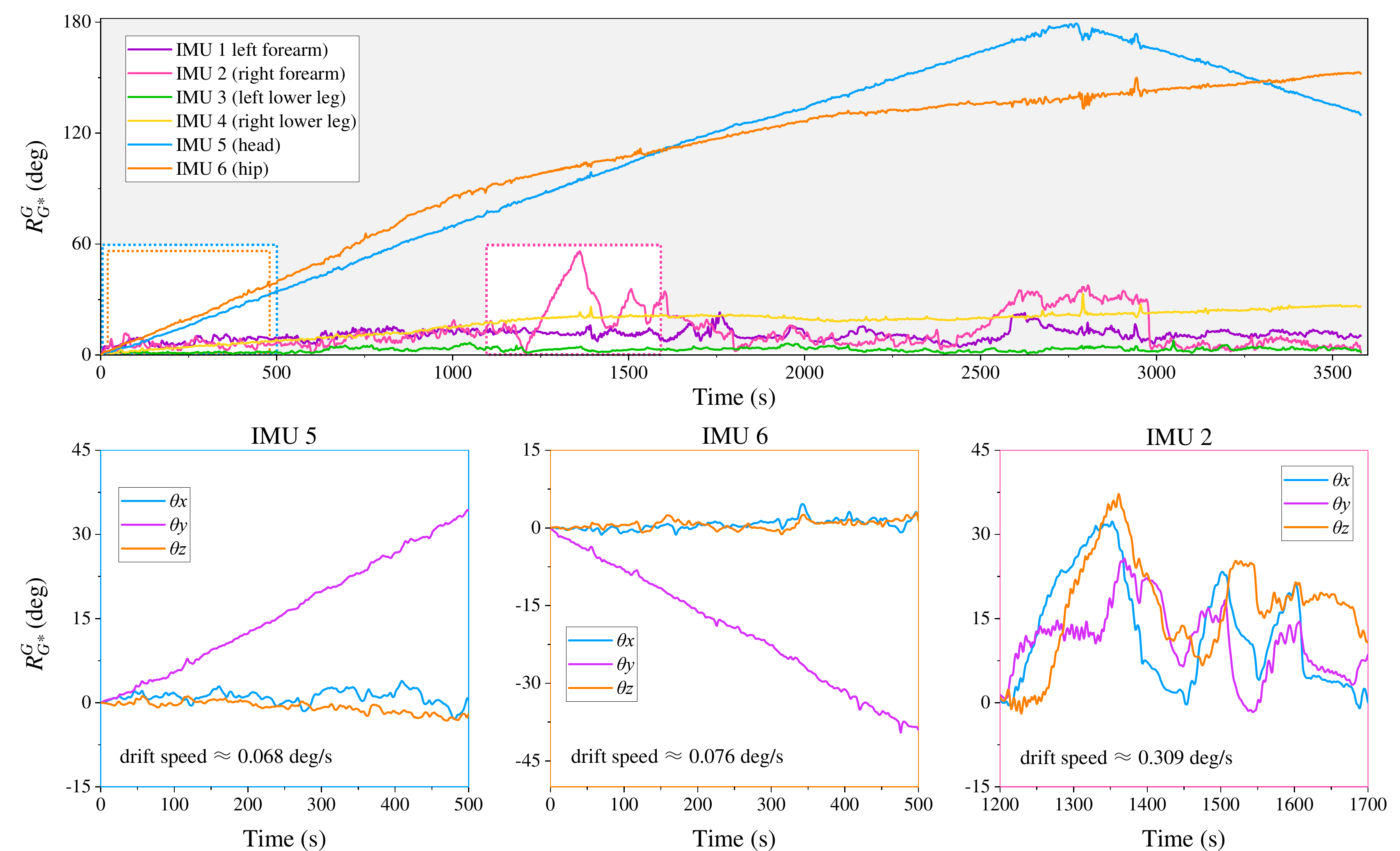}
    \caption{Visualization of IMU drifting in real-world dataset (s1). We use SMPL frame as global frame.}
    \label{fig:imu drift}
\end{figure}

We further verified the IMU drifting when the sensors are placed on human body.
This is achieved by tracking the IMUs, each with 3 optical markers.
This provides the absolute orientation of IMUs, which allows the analysis of IMU drifting.
As shown in Fig.~\ref{fig:imu drift},  $R_{G^{'}G}(t)$ is primarily the y-axis (vertical direction) rotation, while the rotations around the x-axis and z-axis (in the horizontal direction) are smaller. This is because the stable gravity direction can provide a reliable reference for the normal to the horizontal plane, thereby correcting the horizontal drift. However, the vertical rotation relies solely on the magnetometer signal for correction, which is highly susceptible to interference from metal objects, making severe IMU drift more likely to occur.

Meanwhile, we also observed mixed-axis drift in the IMU on the right hand (IMU 2) over short periods, occurring during table tennis movements. This is because during this action, the right hand remains in motion for an extended period, making it difficult to obtain a stable gravity acceleration direction for correcting drift along the x-axis and z-axis.

\section{Calibration Pose Error Analysis}
\begin{figure}[h]
    \centering
    \includegraphics[width=0.49\textwidth]{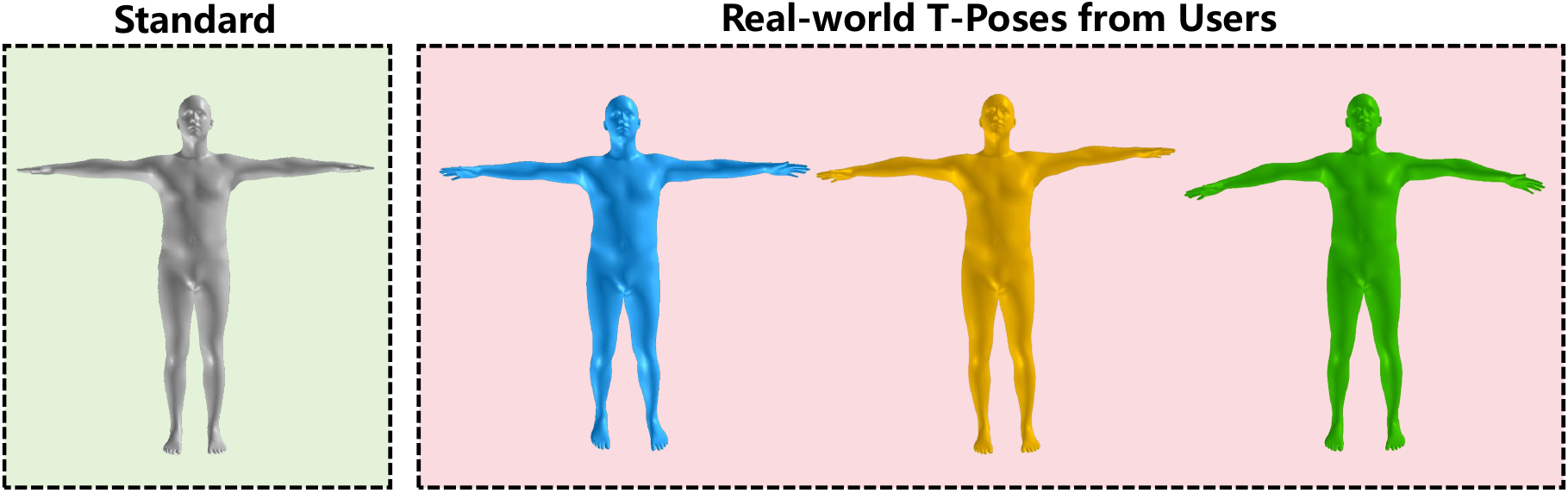}
    \caption{T-Pose error in real-world usage.}
    \label{fig:tpose}
\end{figure}

Static calibration estimates $R_{BS}$ through specific calibration pose (e.g. T-Pose). The drawback of this method is that in real-world usage, the user's T-Pose is difficult to achieve completely accurately (see Fig.~\ref{fig:tpose}), leading to errors in estimating $R_{BS}$, thus affecting the accuracy of pose estimation.

Table~\ref{tab:tpose err} shows the errors in 30 T-Pose instances collected in our real-world dataset, indicating significant errors in the left and right forearms, which were well covered by the distribution of $R_{BS}$ used in our data synthesis (±45° for the x, y and z axes).

\begin{table}[h]
    \centering
    \caption{T-pose err of six joints.}
    \begin{tabular}{lr}
    \toprule
         Joint&  T-Pose Err (deg)\\
        \midrule
         left forearm&  32.86 $\pm$ 11.43\\
         right forearm&29.12 $\pm$ 9.35\\
         left lower leg&11.74 $\pm$ 4.64\\
         right lower leg&13.23 $\pm$ 4.60\\
         head&5.01 $\pm$ 2.57\\
         hip&2.78 $\pm$ 2.19\\
    \bottomrule
    \end{tabular}
    
    \label{tab:tpose err}
\end{table}

\section{Rationale of ego-yaw coordinate system}
\begin{figure}[h]
    \centering
    \includegraphics[width=0.48\textwidth]{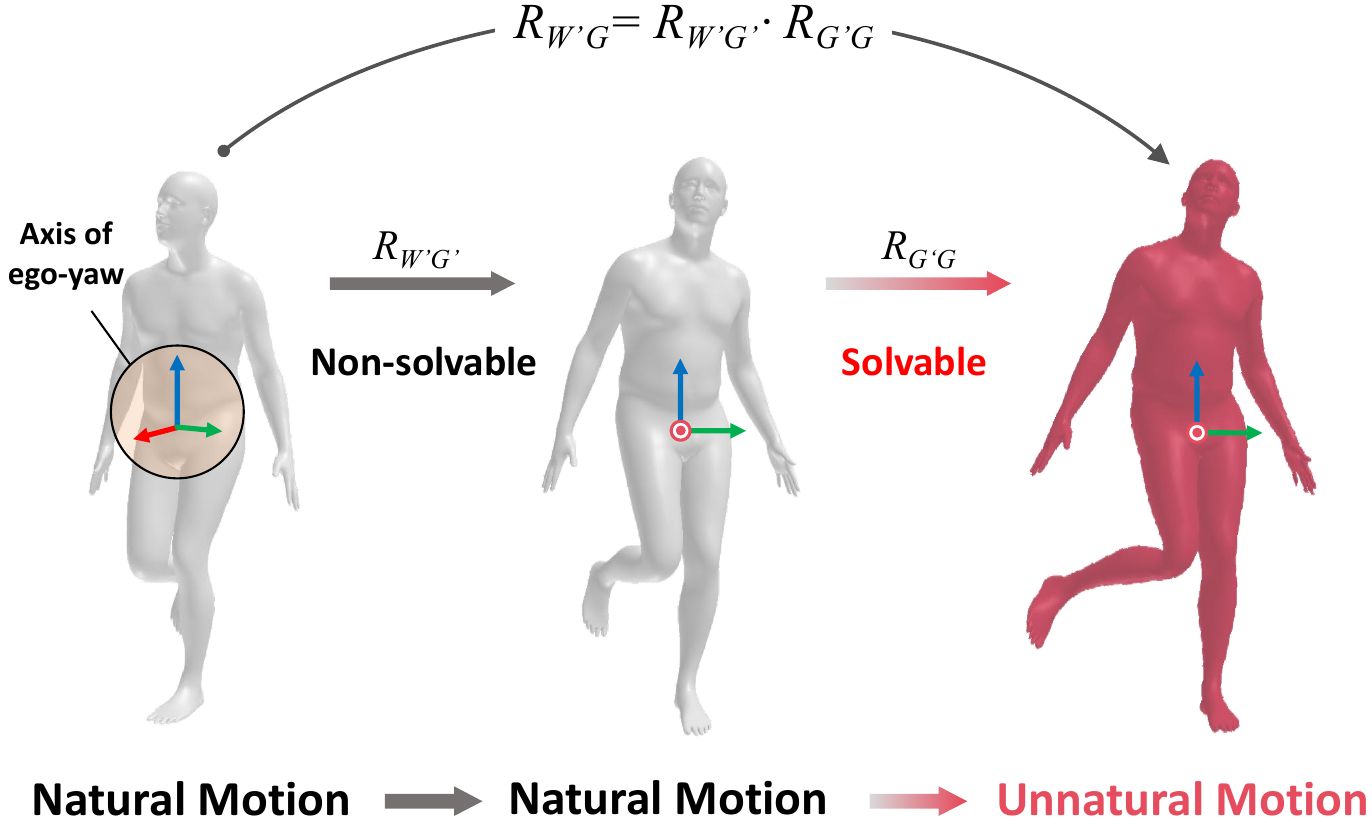}
    \caption{Decompose of the coordinate drift based on the ego-yaw coordinate system.}
    \label{fig:ego-yaw}
\end{figure}

We define the ego-yaw coordinate system as the global coordinate system \( G \) in dynamic calibration due to the exist of unresolved components in the drifted foundational world coordinate systems $W^{'}$ (e.g. ENU or SMPL). As illustrated in Fig.~\ref{fig:ego-yaw}, the overall coordinate drift $R_{W^{'}G}$ can be decomposed into two parts: $R_{W^{'}G^{'}}$, corresponding to yaw rotation between drifted ego-yaw and drifted world coordinate system, which is yaw-only rotation that cannot not result in any unnatural motion; thus, it cannot be perceived and resolved through observation. In contrast, the remaining $R_{G^{'}G}$ directly impacts the rotation of human joints, leading to unnatural movements. Therefore, we set  $R_{G^{'}G}$ as the target of dynamic calibration.


\section{Hyperparameters Selection}
\subsection{$T_R$ in Calibration Trigger}
To ensure accurate calibration parameter estimation, setting a high $T_R$ is a safe strategy. However, since everyday activities do not always maintain high $RD$, this could lead to calibration not being triggered in a timely manner to track changes in calibration parameters. Therefore, the $T_R$ setting needs to balance calibration accuracy and trigger frequency.

To obtain the reference data for the $T_R$ settings, we conducted experiments on the KIT~\cite{mandery2015kit} subset of the AMASS dataset. 
The KIT dataset contains a large number of everyday actions, and we recorded the average rotation diversity of each consecutive $\hat{R}_{IMU}$ sequence sample per 512 frames (60Hz), as well as the OME of TIC on its corresponding 2000 randomly synthesized samples. 

\begin{figure}[h]
    \centering
    \includegraphics[width=0.48\textwidth]{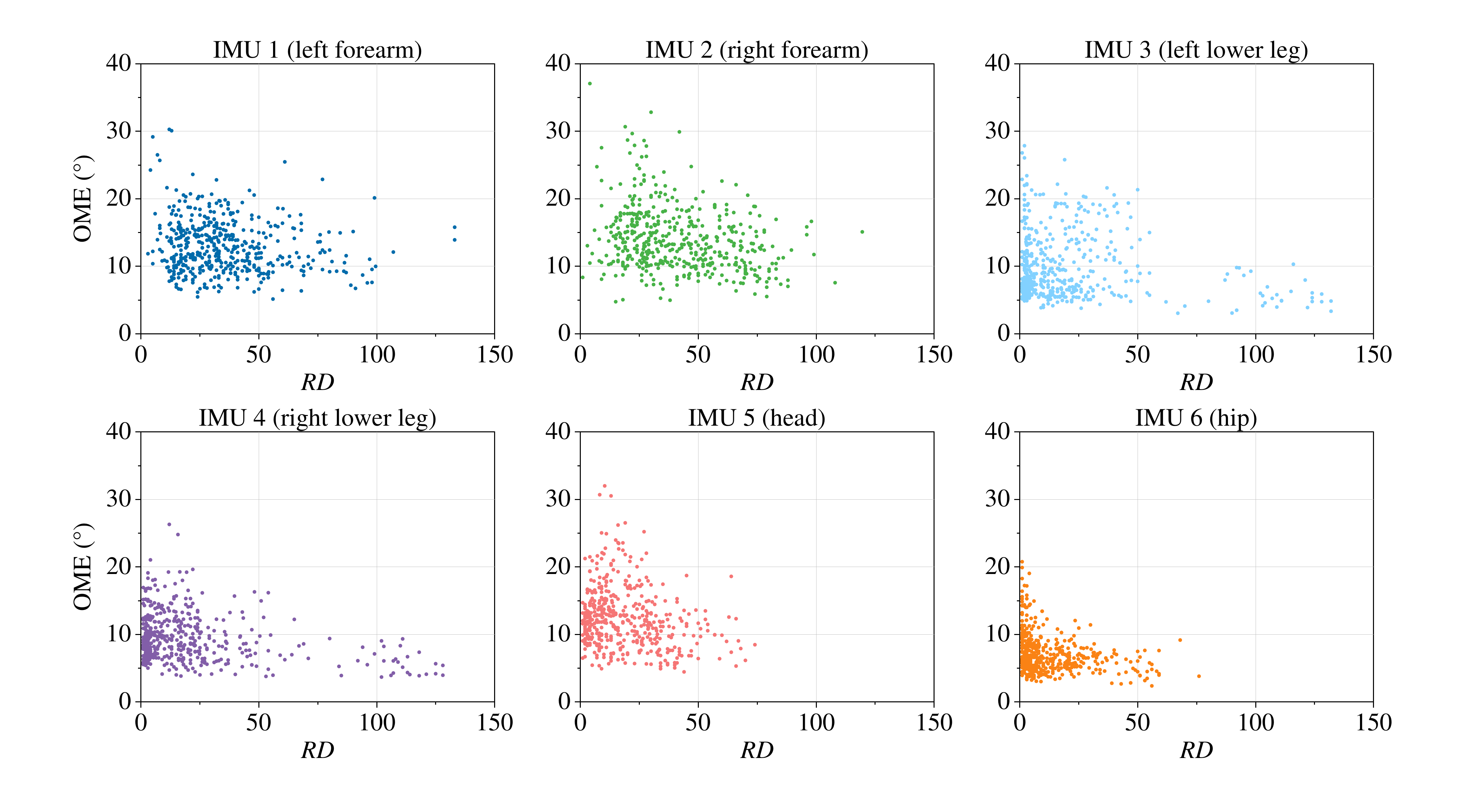}
    \vspace{-4mm}
    \caption{The impact of rotation diversity on OME.}
    \vspace{-2mm}
    \label{fig:rd ana}
\end{figure}

As shown in Fig.~\ref{fig:rd ana}, we can observe that for each joint, lower RD tends to lead to higher OME. , demonstrating the importance of satisfying Assum.3 for accurate calibration parameter estimation. We also noticed that different joints have different $T_{R}$ requirements. 

In Table.~\ref{tab:rd ana}, we present the final $T_R$ we used, along with three indexes that guide the $T_R$ settings for each joint: 1) $\overline{RD}$, the average $RD$ across all motion samples; 2) $\overline{RD}_{<10}$, the average $RD$ of samples where OME < 10° across all motion samples; and 3) $s_{RD}$, the $RD$ sensitivity, defined as $s_{RD}=\overline{RD}_{<10} / \overline{RD}$, which represents the sensitivity of OME to $RD$. For joints with a higher $s_{RD}$, we tend to select $T_R$ values above $\overline{RD}$ to ensure accuracy; conversely, for joints with a lower $s_{RD}$, we prefer $T_R$ values below $\overline{RD}$ to ensure trigger frequency.

\begin{table}[h]
    \centering
     \caption{$T_R$ setting in TIC. $\overline{RD}$ and
     $\overline{RD}_{<10}$ are calculated from samples in Fig.~\ref{fig:rd ana}.}
    \begin{tabular}{lcccc}
    \toprule
         Joint &$\overline{RD}$&  $\overline{RD}_{<10}$&$s_{RD}$&  $T_R$\\
         \midrule
         left forearm &35.59&  38.04 &1.07& 30\\
         right forearm &41.85&  56.99 &1.36& 50\\
         left lower leg
 &27.46& 30.96 &1.13&30\\
         right lower leg &27.01&  32.20 &1.19& 30\\
         head &22.09&  28.40 &1.29& 25\\
         hip &17.03&  18.50 &1.08& 15\\
    \bottomrule
    \end{tabular}
    \label{tab:rd ana}
\end{table}

\begin{figure}
    \centering
    \includegraphics[width=0.40\textwidth]{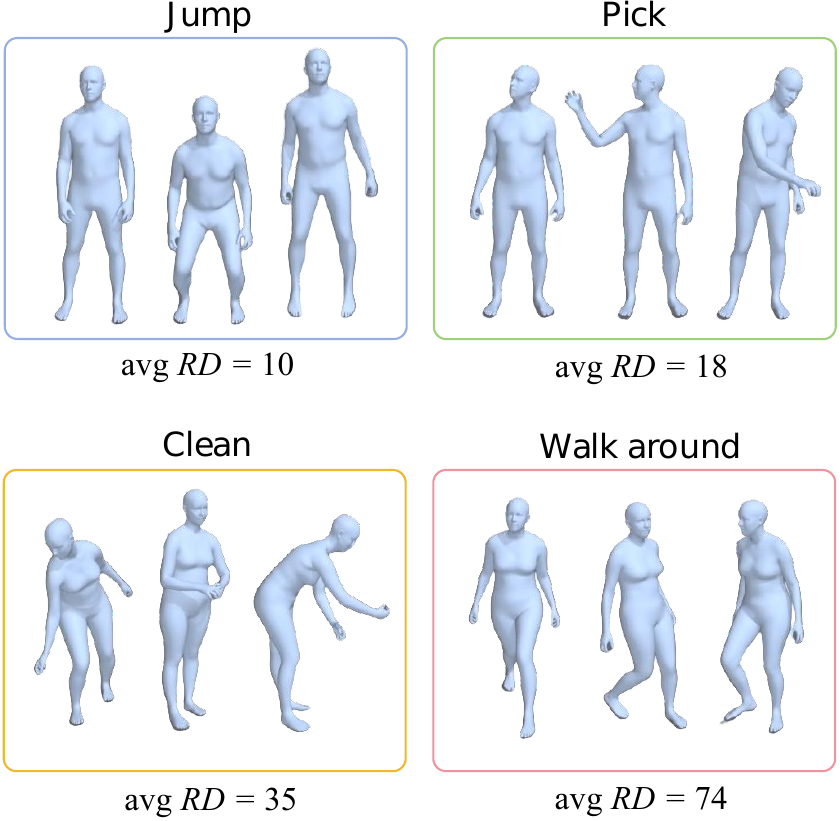}
    \caption{Four examples of daily activities from KIT dataset. Motion labels and visualization are provided by BABEL dataset~\cite{punnakkal2021babel}.}
    \label{fig:pose rd}
\end{figure}

Fig~\ref{fig:pose rd} shows the average $RD$ for different daily activities. We found that activities involving body movement (such as cleaning, walking around) tend to cause changes in body's facing detection, resulting in an average $RD$ greater than 30. Thus, our $T_R$ settings  (avg=30) is appropriate and ensures that  dynamic calibration can be triggered in time during daily use.

\subsection{Time intervals $t$ of timing signal in dynamic calibration}
During the inertial motion capture process, we use a timing signal $S_t$ with an interval of $t$to run the TIC network. Because in general, the changing of the calibration parameters is slow (e.g., coordinate drift of $0.1$ deg/s), allowing for updates at a lower frequency to avoid unnecessary computational cost. We tested the calibration accuracy under different $t$ settings. As shown in the table~\ref{tab:time intervals settings}, when $t \leq 2$ seconds, there is no significant difference between OME and AME. However, as $t$ increases, OME begins to rise, and a notable decline in performance is observed in both OME and AME at $t = 20$ seconds. Consequently, we have chosen $t = 1$ seconds to achieve a trade-off between performance and computational cost.
\begin{table}
    \centering
    \caption{TIC performance under different running time interval $t$.}
    \begin{tabular}{ccc}
    \toprule
         $t$ (sec)&  Avg OME (deg)& Avg AME  ($m/s^2$)\\
         \midrule
         0.5&  15.20& 1.30\\
         1&  15.20& 1.30\\
         2&  15.39& 1.31\\
         5&  15.42& 1.31\\
         10& 15.54&1.31\\
         20& 15.85&1.34\\
 \bottomrule
    \end{tabular}
    
    \label{tab:time intervals settings}
\end{table}

\subsection{Resolution of discretized Euler angle space}
The ideal discretized Euler angle space should use the lowest possible split step to achieve higher resolution and accurately represent rotation diversity. However, since the Euler angle space is three-dimensional, too high a resolution would greatly increase the computational complexity of $RD$ calculation. Therefore, we use a 15-degree split step for discretization, which ensures uniform segmentation with an acceptable $RD$ computation complexity (takes 2.5ms to process 256 frames data on an NVDIA RTX 4080 GPU).

\section{TIC Network}
\subsection{Network Architecture}
The transformer encoder blocks in Encoder (E) of TIC network and TPM module use the same network architecture, with details shown in Table~\ref{tab:network details}.
\begin{table}[h]
    \centering
     \caption{Network details of transformer encoder blocks  in Encoder (E) of TIC network and TPM module.}
    \begin{tabular}{lr}
    \toprule
         Param&  Value\\
        \midrule
         Embedding Dim&256\\
         Attention Heads&8\\
         FFN Size&512\\
    \bottomrule
    \end{tabular}
    \label{tab:network details}
    \vspace{-5mm}
\end{table}
\revision{
\subsection{Choice of Architecture}
Our TIC network features three key designs as follows:
\begin{itemize}
    \item Transformer-backbone: The Transformer architecture was chosen for its advantages in sequence modeling.
    \item Encoder-only: The predictions of $\Delta R_{G^{'}G}$ and $\Delta R_{BS}$ are fixed-size outputs rather than sequences.
    \item Temporal Average Pooling: This design supports variable input lengths, allowing our TIC network to adapt to different motion speeds or sampling rates without retraining.
\end{itemize}
}
\begin{figure}[h]
    \centering
    \includegraphics[width=0.40\textwidth]{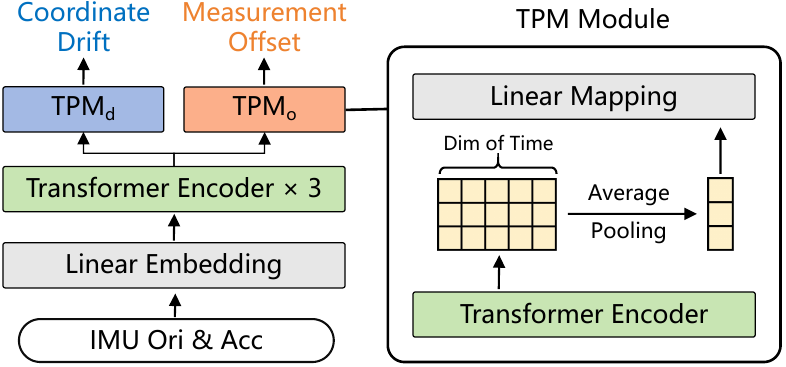}
    \caption{Architecture of TIC network. The network uses three stacked Transformer Encoders to extract features from IMU orientation and acceleration sequences, and utilizes two TPM (Transformer Encoder-Pooling-Mapping) modules to estimate coordinate drift and measurement bias, respectively.}
    \label{fig:tic network}
\end{figure}
\section{Training Details}

\paragraph{Data Format}
In model training, the input of the TIC Network are sequences of concatenated acceleration $a_{\rm IMU}\in \mathbb{R}^{n \times (6\times3)}$ and orientation (rotation matrices) $R_{\rm IMU}\in \mathbb{R}^{n \times (6\times3\times3)}$ from 6 IMUs, where $n=256$, indicating the length of the sequence. The $a_{\rm IMU}$ were divide by 30. The Ground Truth $R_{G^{'}G}\in \mathbb{R}^{n \times (6\times6)}$, $R_{BS}\in \mathbb{R}^{n \times (6\times6)}$ were converted into 6D representations~\cite{zhou2019continuity}, .

\paragraph{Training Settings}
All experiments were conducted on a computer equipped with an Intel(R) Core(TM) i7-13700KF CPU and an NVIDIA RTX 4080 GPU. The TIC network was implemented using PyTorch 1.12.1 with CUDA 11.3. The training batch size was set to 128, learning rate to 0.001, and the Adam optimizer with default parameters was used. 

The training process includes two steps:   
1) \textit{Pre-training.} The model is trained for 10 epochs using $\text{DS}_{\text{AMS}}$.   
2) \textit{Fine-tuning.} The model is trained for 3 epoch using the concatenated datasets $\text{DS}_{\text{AMS}}$ and $\text{DS}_{\text{DIP}}$.


\begin{table}
    \centering
    \caption{Reproducibility analysis of TIC Network with 10 independent model training. Min/Max OME: 14.96/15.79; Min/Max AME: 1.30/1.32.}
    \begin{tabular}{ccc}
    \toprule
         Joint&  OME (deg)& AME ($m/s^2$)\\
         \midrule
         left forearm
&  21.35 $\pm$ 0.95& 1.52 $\pm$ 0.03\\
         right forearm
&  20.52 $\pm$ 1.06& 1.90 $\pm$ 0.05\\
         left lower leg
&  13.63 $\pm$ 0.37& 1.54 $\pm$ 0.01\\
         right lower leg
&  12.92 $\pm$ 0.39& 1.42 $\pm$ 0.01\\
         head
&  16.51 $\pm$ 0.48& 0.63 $\pm$ 0.02\\
         hip&  6.04 $\pm$ 0.44& 0.81 $\pm$ 0.01\\
         \midrule
         Avg&  15.17 $\pm$ 0.23& 1.30 $\pm$ 0.01\\
         \bottomrule
    \end{tabular}
    
    \label{tab:reproducibility}
\end{table}

\section{Deployment}
\subsection{Re-calibration Cost}
With an NVIDIA RTX 4080 GPU, the re-calibration requires only 4.5ms (2.5ms for RD computation, 2ms for TIC network inference), yielding an FPS of 222.2, which fully meets the real-time requirement.

\subsection{Calibration Time w.r.t Initial Calibration Error}
As Table~\ref{tab:calibration timecost} shows, we selected 21 samples with significant initial calibration errors from $\text{DS}_{\text{TIC}}$ (Avg OME > 15°) and recorded the time costs to complete the calibration. It can be observed that average time costs grow with initial calibration errors. Nevertheless, the min time costs (last column) indicate that the calibration can be significantly accelerated by performing high-RD movements to enable rapid triggering and accurate $R_{G^{'}G}$ and $R_{BS}$ estimation.
\begin{table}[h]  
\centering
\caption{\revision{Statistic of Calibration Time w.r.t Initial Calibration Error.}}
\begin{tabular}{crr}
\toprule
Initial OME (deg) & Avg Time Cost (s) & Min Time Cost (s)\\
\midrule 
15-30  & 20.65$\pm$23.24 & 7.26 (Avg $RD$=19.17) \\
30-60 & 40.05$\pm$21.92 & 16.48 (Avg $RD$=24.50) \\
60-100  & 128.91$\pm$93.01 & 10.28 (Avg $RD$=28.16) \\
\bottomrule
\end{tabular}
\label{tab:calibration timecost}
\end{table}  

\section{Evaluation on Xsens IMU}
\begin{figure}[t]
    \centering
    \includegraphics[width=0.40\textwidth]{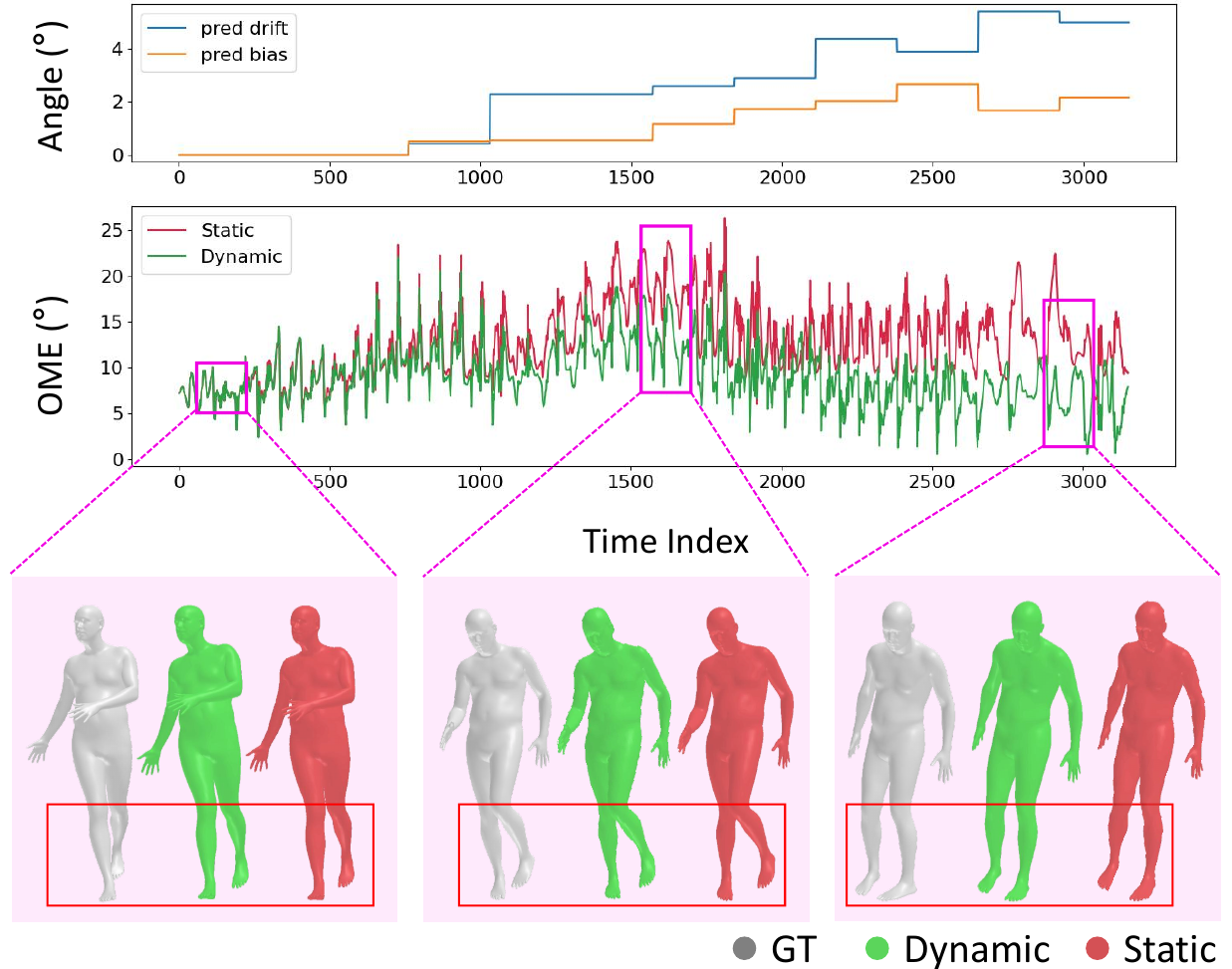}
    \caption{An example of our dynamic calibration on Xsens IMUs. We visualize the orientation measurement error (OME) along with the predicted calibration parameters, which include the rotation angle of $R_{G^{'}G}$ and $R_{BS}$ of the right lower leg.}
    \label{fig:total capture vis}
\end{figure}

We manually selected a data segment with a high OME (8.86°) from the Total Capture dataset~\cite{trumble2017total} (collected with Xsens IMUs) and applied TIC for dynamic calibration. As shown in Table \ref{tab:xsens}, although there is a slight increase in OME for the right forearm, head, and waist, the overall OME still decreased as expected (8.86°$\rightarrow$7.92°). Fig.~\ref{fig:total capture vis} illustrates the process of OME reduction of the right lower leg using our dynamic calibration. After the motion begins, the OME decreases with the updates of the coordinate drift $R_{G^{'}G}$ and measurement bias $R_{BS}$. This demonstrates that our TIC does not depend on specific IMU device and has the potential for application in different types of inertial motion capture systems.
\begin{table}
    \centering
    \caption{Evaluation on Xsens IMUs. The data are sampled from Total Capture Dataset}
    \begin{tabular}{ccc}
    \toprule
         \multirow{2}{*}{Joint}&  \multicolumn{2}{c}{OME (deg)}\\
         \cline{2-3}
 & with TIC&without TIC\\
         \midrule
         left forearm
&  \textbf{8.02}& 10.93\\
         right forearm
&  7.67& \textbf{7.37}\\
         left lower leg
&  \textbf{13.79}& 14.19\\
         right lower leg
&  \textbf{9.13}& 12.71\\
         head
&  5.58& \textbf{4.79}\\
         hip&  3.34& \textbf{3.15}\\
         \midrule
         Avg&  \textbf{7.92}& 8.86\\
         \bottomrule
    \end{tabular}
    
    \label{tab:xsens}
\end{table}



\section{Reproducibility Test}
The use of randomly generated samples $R_{G^{'}G}$ and $R_{BS}$ in TIC Network training introduces additional uncertainty into the model training process, making reproducibility analysis essential. We performed 10 independent trainings of the TIC Network under a unified training setup and recorded the average OME and AME. As shown in Table~\ref{tab:reproducibility}, the performance variations across the 10 independent model trainings were minimal, indicating that the achieved metrics (OME: 15.20, AME: 1.30) are highly reproducible.

\revision{
\section{Choice of $\Delta R_{G^{'}G}$ \& $\Delta R_{BS}$ v.s. $R_{G^{'}G}$ \& $R_{BS}$}
We use the differences $\Delta R_{G^{'}G}$ \& $\Delta R_{BS}$ as they yield smaller output variation than $R_{G^{'}G}$ \& $R_{BS}$, thereby facilitating calibration accuracy as shown in Table~\ref{tab:differences estimation}.
}

\begin{table}[h]  
\centering  
\caption{\revision{Comparison of differences update and absolute value update.}}  
\begin{tabular}{lrrrr}  
\toprule  
Method & OME & AME & $R_{G^{'}G}$ Err & $R_{BS}$ Err \\
\midrule  
$\Delta R_{G^{'}G}$ \& $\Delta R_{BS}$ & 15.20& 1.30& 8.41 & 15.79\\
$R_{G^{'}G}$ \& $R_{BS}$ & 17.34 & 1.40 & 8.79 & 17.08 \\
\bottomrule  
\end{tabular}
\label{tab:differences estimation}
\end{table}

\section{TIC v.s. End-to-End Regression}
The proposed TIC ensures accurate measurement of skeletal orientation and global acceleration through dynamic calibration parameter updates. However, a naive implementation of dynamic calibration is to use the TIC network to directly estimate the already calibrated data, known as End-to-End (End2End) Regression. To compare the effectiveness of these two approaches, we replaced the TPM module of the TIC Network with a DNN, used to estimate the calibrated orientation and acceleration. The modified model employed the same training settings as the TIC Network and was tested on real datasets. 

\begin{table}[h]
    \caption{Competition of TIC and End-to-End regression.}
    \centering
    \begin{tabular}{lcccc}
    \toprule
        \multirow{2}{*}{Joint}&  \multicolumn{2}{c}{OME (deg)} & \multicolumn{2}{c}{AME ($m/s^2$)}\\
        \cline{2-5}
                      &  TIC& End2End&  TIC& End2End\\
        \midrule
         left forearm &  {\bf21.40}
&   24.68& {\bf1.53}
&1.91\\
         right forearm&  {\bf20.56}
&  26.42& {\bf1.90}
&2.84\\
         left lower leg&  {\bf13.79}
&  14.69& {\bf1.54}
&2.11\\
         right lower leg&  {\bf12.93}
& 15.66& {\bf1.42}
&2.00\\
         head          &  {\bf16.56}
&  16.92& {\bf0.62}
&1.25\\
         hip           & {\bf6.00}
& 6.40& {\bf0.80}
&1.16\\
         \midrule
         Avg           & {\bf15.20}&17.46& {\bf1.30}&1.88\\
    \bottomrule
    \end{tabular}
    
    \label{tab:tic vs end2end}
\end{table}
As shown in Table~\ref{tab:tic vs end2end}, the OME and AME of End2End Regression are noticeably higher than those of TIC. This is because: \textbf{1) Cannot guarantee the satisfaction of ASSUM 3.} End2End Regression must operate in sync with motion capture at the same frame rate to provide real-time calibration, even when ASSUM 3 is not met. \textbf{2) Lack of prior regularization.} IMU calibration is a well-formulated process based on calibration parameters ($R_{G^{'}G}$ and $R_{BS}$), and End2End Regression cannot incorporate such prior knowledge, leading to poor generalization . For example, acceleration calibration under arbitrary human motion can be accomplished via known $R_{G^{'}G}$ and the acceleration measurement modeling (eq.~\ref{eq:real acc drift}), but using End2End Regression would require the training data to include all possible human motions, which is challenging to satisfy.

\end{document}